\documentclass[%
 reprint,
 amsmath,amssymb,
 aps,
]{revtex4-1}

 \usepackage[usenames,dvipsnames]{pstricks}
 \usepackage{epsfig}
\usepackage{pst-grad} 
\usepackage{pst-plot} 
\usepackage{graphicx}
\usepackage{dcolumn}
\usepackage{bm}
\usepackage{subfigure}  

\newcommand{\bs}[1]{\boldsymbol{#1}}

\newcommand{\A}[1]{\Hat{#1}}

\newcommand{\T}[1]{\widetilde{#1}}
\newcommand{\pd}[2]{\dfrac{\partial #1}{\partial #2}}

\newcommand{\pds}[1]{\dfrac{\partial}{\partial #1}}

\newcommand{\pdss}[1]{\dfrac{\partial^2}{\partial {#1}^2}}

\newcommand{\pdt}[2]{\dfrac{\partial^2 {#1}}{\partial {#2}^2}}

\newcommand{\vp}[1]{\mathbf{#1}}

\newcommand{\abs}[1]{\left| #1\right|}
\newcommand{\ket}[1]{|#1\rangle}
\newcommand{\bra}[1]{\langle #1| }

\newcommand{\Ex}[1]{\langle #1\rangle}
\DeclareMathOperator{\Tr}{Tr}
\newcommand{\bo}[1]{\boldsymbol{#1}}
\newcommand{\OO}[1]{\overline{\overline{#1}}}

\begin{document}

\preprint{APS/123-QED}

\title{Nonlinear quantum optical properties of graphene: the role of chirality and symmetry}

\author{Behrooz Semnani $^{1,2}$}
 \email{bsemnani@uwaterloo.ca}
\author{Amir Hamed Majedi $^{1,2,3,4}$}%
 \email{ahmajedi@uwaterloo.ca}

\author{Safieddin Safavi-Naeini $^{1}$}
 \email{safavi@uwaterloo.ca}
\affiliation{%
 $^1$ Department of Electrical \& Computer Engineering, University of Waterloo, Waterloo, ON, Canada\\
 $^{2}$ Waterloo Institute for Nanotechnology, University of Waterloo, Waterloo, ON, Canada\\
 $^{3}$ Department of Physics \& Astronomy , University of Waterloo, Waterloo, ON,  Canada\\
 $^{4}$ Perimeter Institute for the Theoretical Physics (PI) , Waterloo, ON, Canada
}%

\date{\today}

\begin{abstract}
We present a semiclassical theory of linear and nonlinear optical response of graphene.
The emphasis is placed on the nonlinear optical response of graphene from the standpoint of the underlying chiral symmetry. The Bloch quasiparticles in low energy limit, around the degeneracy points are dominantly chiral. It is shown for the first time that this chiral behavior in conjunction with scale invariance in graphene around the Dirac points results in the strong nonlinear optical response. Explicit expressions for the linear and nonlinear conductivity tensors are derived based on Semiconductor Bloch Equations (SBEs). The linear terms agree with the result of Kubo formulation. The three main additive mechanisms contribute in the nonlinear optical response of graphene: pure intraband, pure interband and the interplay between them. For each contribution, an explicit response function is derived. The Kerr-type nonlinearity of graphene is then studied and it is demonstrated that its Kerr nonlinear coefficient is several orders of magnitude higher than that of many other known semiconductors. In addition, the nonlinear refractive index of graphene can also be tuned and enhanced by applying a gate voltage.

\end{abstract}

\maketitle

\section{Introduction}
Graphene is a  two dimensional arrangement of the carbon atoms sitting in a hexagonal lattice, a seemingly simple lattice structure that nonetheless underlies the special transport and optical properties \cite{neto2009electronic}. The band structure of graphene differs substantially from other condensed matter systems. The effective Hamiltonian describes pseudo-relativistic quasiparticles obeying $(2+1)$-dimensional Dirac equation. In the context  of  QED, the electronic excitations introduced by such dynamics can be considered as massless chiral fermions \cite{novoselov2005two}.

Graphene exhibits a variety of peculiar properties that are manifestations of the special symmetries of its crystalline structure and relativistic energy spectrum of charged carriers. Symmetries entail several unconventional properties such as the existence of a topologically protected zero-energy state, Berry phase, anomalous quantum Hall effect and Zitterbewegung (`trembling motion') \cite{neto2009electronic,zhang2005experimental,katsnelson2006chiral}. It is counterintuitive that intrinsic graphene has a finite conductivity, in the order of the Hall conductivity $e^2/\hbar$, at zero temperature and zero carrier concentration. The current operator does not commute with the Hamiltonian and therefore, graphene cannot sustain the current. This \textit{intrisic disorder} leads to a finite conductivity \cite{katsnelson}. All these odd properties can be linked to the chiral behavior of the carriers. In graphene the pseudospin is locked parallel or antiparallel to the direction along which the electron propagates and so the quasiparticles possess the property of chirality \cite{katsnelson}.

The optical response of graphene is also expected to be influenced by the chiral nature of the carriers and the scale invariance of the band structure in low energy limit. However, despite its importance, a theoretical study on the unconventional optical response of graphene is still lacking. The optical response of graphene in the linear regime has been investigated theoretically and experimentally 
\cite{gusynin2007magneto,gusynin2007ac,hwang2007dielectric,stauber2008optical,horng2011drude,dawlaty2008measurement}. 
Graphene as a scale invariant two dimensional chiral electronic system exhibits universal optical response \cite{bacsi2013}. A simple analysis based on linear response theory shows that an isolated graphene sheet can absorb about 2.3\% of the normally incident optical field, which is indeed a huge number for a monolayer atomic structure. The nonlinear optical response of graphene has been a topic of intensive research in the recent years \cite{mikhailov2008nonlinear,wright2009strong,ishikawa2010nonlinear,roberts2011response,hong2013optical,ishikawa2013electronic,cheng2014third,mikhailov2014quantum,Jafari2012,Cheng2014}. Treatment of the nonlinear optical response of graphene in the framework of quasiclassical Boltzman equation predicts strong nonlinearity in the terahertz range of frequency, neglecting pair productions and interband transitions \cite{mikhailov2008nonlinear}. This part of nonlinearity is mainly due to the geometrical properties of band structure rather than its topological aspects \cite{hatsugai2011topological}. The calculation of optical response of graphene in time domain has been carried out in Ref.~\cite{ishikawa2010nonlinear}. Wrighte \textit{et al.} have performed Fourier analysis of the dirac equation to obtain the optical response of the system for a given incident field \cite{wright2009strong}. All existing  time domain methods proceed primarily at the level of the wavefunction, rather than at the level of the density matrix
and thereby suffer from the computational cost and the difficulty resulted from the inclusion of relaxation processes due to impurities and emission. The nonlinear optical resposen in gapped graphene, where the low-energy single-particle spectrum is modeled by massive Dirac equation is discussed in \cite{Jafari2012}.

The treatment of optical response of graphene, in a simplistic manner, proceeds from single particle approximation and the equation of motion for density matrix. The formal approach to calculate the optical response of semiconductors, excited by a electromagnetic field, is based on perturbative expansion of the density matrix and taking all possible transitions into account \cite{boyd2003nonlinear,mikhailov2014quantum}. Even at such a reasonably simplistic treatment, the evaluation of nonlinear response coefficients is numerically difficult and does not provide any intuitive insight.  To circumvent these difficulties, Semiconductor Bloch Equations (SBEs) for graphene are employed.

SBEs \cite{klingshirn2012semiconductor} have extensively been used to calculate the nonlinear optical response of semiconductors \cite{lindberg1988effective,aversa1995nonlinear}. In Ref.~\cite{avetissian2013nonlinear} the derivation of SBEs for graphene beyond the Dirac cone approximation has been discussed. In Ref.~ \cite{stroucken2011optical} the general treatment of the SBEs for graphene including electron-electron interactions and exciton effects is presented. Using SBEs, the problem of interaction can be treated in a semiclassical manner leading to  numerically amenable expressions for arbitrary orders of interaction. SBEs introduce an effective dipole in the reciprocal space revealing peculiarities of graphene in terms of its optical response. We have shown that, in the absence of spin interactions, the aforementioned dipole is singular at the high symmetry points of the reciprocal space  leading to a strong nonlinear optical response.

In this paper we study the optical response of graphene based on a semicassical theory. We will adopt an approach that treats the electrons dynamics in the presence of a moderate intensity electromagnetic field based on SBEs. We demonstrate that the higher order nonlinear response of graphene, possesses a singularity due to the topological properties of the band structure and the chiral nature of the charged carriers. To remove this singularity the inclusion of the spin-orbit interaction in a phenomenological level  is proposed.

The paper is organized as follows. In Sec.~\ref{section:2} we present the Hamiltonian of graphene within the tight binding approximation. In Sec.~\ref{section:3} we address the question of how the chirality of carriers affects the optical response of graphene and its dependence on the fundamental group symmetries of the problem. In Sec.~ \ref{section:4} the equations of motion for the single particle density matrix are formulated and the SBEs are derived in the context of Dirac equation. In Sec.~\ref{section:5} we propose an iterative approach to solve the effective optical Bloch equations. Secs.~\ref{section:6} and \ref{section:7} present the derivation of linear and third order nonlinear conductivities. The numerical results and discussions on the importance of graphene as a strong nonlinear material are given in Sec.~\ref{section:8}. We will show that nonlinear effects in graphene are substantially stronger than those of other known semiconductors. In Sec.~ \ref{section:conclusion} we summarize our results.

\section{Graphene Hamiltonian and Equations of Motion \label{section:2}}
Graphene has a honeycomb crystal lattice with two lattice points per elementary cell. They belong to two sublattices A and B where the nearest neighbours of the sites of one of them are sites belonging to the other sublattice (a bipartite lattice). In Fig.~\ref{fig_graphene_a} atoms in A and B sublattices are shown by blue and red balls respectively. The Bravais lattice is triangular with the lattice vectors given by \cite{neto2009electronic},
\begin{equation}
\vp{a}_1=a\left(\frac{3}{2}\A{x}+\frac{\sqrt{3}}{2}\A{y}\right)\quad,\quad \vp{a}_2=a\left(\frac{3}{2}\A{x}-\frac{\sqrt{3}}{2}\A{y}\right)
\end{equation} 
As shown in Fig.~\ref{fig_graphene_b} the reciprocal lattice is also hexagonal with rhomboidal unit cell formed by two vectors
\begin{equation}
\vp{b}_1=\frac{2\pi}{3a}\left(\A{x}+\sqrt{3}\A{y}\right)\quad ,\quad \vp{b}_2=\frac{2\pi}{3a}\left(\A{x}-\sqrt{3}\A{y}\right)
\end{equation}
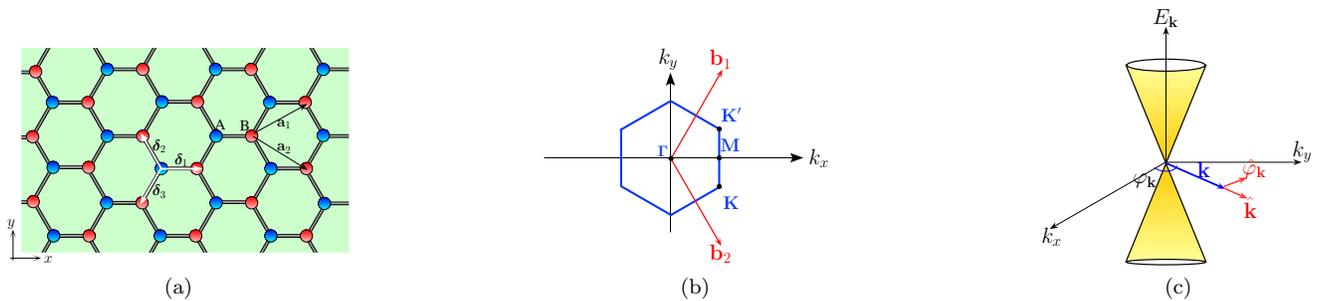
\begin{figure*}
    \centering
    \begin{subfigure}[]{
        \centering
    
\scalebox{0.4} 
{
\begin{pspicture}(0,-3.7367187)(11.682656,3.6967187)
\definecolor{color717g}{rgb}{1.0,0.8,0.8}
\definecolor{color2430g}{rgb}{0.8,0.8,0.8}
\definecolor{color2430f}{rgb}{0.4,0.4,0.4}
\definecolor{color322b}{rgb}{0.8,1.0,0.8}
\psframe[linewidth=0.002,linecolor=white,dimen=outer,fillstyle=solid,fillcolor=color322b](11.642189,3.5967188)(0.5421875,-3.4632814)
\rput{-120.0}(6.4651937,6.0261197){\psframe[linewidth=0.01,dimen=outer,fillstyle=gradient,gradlines=2000,gradbegin=color2430g,gradend=color2430f,gradmidpoint=1.0](5.5821877,1.196719)(4.362188,1.096719)}
\rput{120.0}(11.211368,-4.179449){\psframe[linewidth=0.01,dimen=outer,fillstyle=gradient,gradlines=2000,gradbegin=color2430g,gradend=color2430f,gradmidpoint=1.0](7.422187,1.1967185)(6.2021866,1.0967184)}
\rput{-120.0}(8.229887,9.204966){\psframe[linewidth=0.01,dimen=outer,fillstyle=gradient,gradlines=2000,gradbegin=color2430g,gradend=color2430f,gradmidpoint=1.0](7.3821883,2.276719)(6.162188,2.1767192)}
\rput{120.0}(9.403996,-0.93596274){\psframe[linewidth=0.01,dimen=outer,fillstyle=gradient,gradlines=2000,gradbegin=color2430g,gradend=color2430f,gradmidpoint=1.0](5.5821877,2.2967188)(4.362188,2.1967187)}
\psframe[linewidth=0.01,dimen=outer,fillstyle=gradient,gradlines=2000,gradbegin=color2430g,gradend=color2430f,gradmidpoint=1.0](6.5021873,1.7567188)(5.282188,1.6567187)
\pscircle[linewidth=0.012,dimen=outer,fillstyle=gradient,gradlines=2000,gradmidpoint=1.0](5.292187,1.7067188){0.21}
\pscircle[linewidth=0.012,dimen=outer,fillstyle=gradient,gradlines=2000,gradbegin=red,gradend=color717g,gradmidpoint=1.0](6.472187,1.7067188){0.21}
\rput{120.0}(7.4641004,-4.295962){\psframe[linewidth=0.01,dimen=outer,fillstyle=gradient,gradlines=2000,gradbegin=color2430g,gradend=color2430f,gradmidpoint=1.0](5.582187,0.056719214)(4.362188,-0.043280788)}
\rput{-120.0}(4.7478213,2.8345938){\psframe[linewidth=0.01,dimen=outer,fillstyle=gradient,gradlines=2000,gradbegin=color2430g,gradend=color2430f,gradmidpoint=1.0](3.8021874,0.0967189)(2.5821874,-0.003281101)}
\rput{120.0}(5.656728,-1.0524759){\psframe[linewidth=0.01,dimen=outer,fillstyle=gradient,gradlines=2000,gradbegin=color2430g,gradend=color2430f,gradmidpoint=1.0](3.7421875,1.1567187)(2.5221875,1.0567187)}
\psframe[linewidth=0.01,dimen=outer,fillstyle=gradient,gradlines=2000,gradbegin=color2430g,gradend=color2430f,gradmidpoint=1.0](4.682187,0.61671877)(3.4621875,0.5167187)
\pscircle[linewidth=0.012,dimen=outer,fillstyle=gradient,gradlines=2000,gradmidpoint=1.0](3.4721875,0.56671876){0.21}
\pscircle[linewidth=0.012,dimen=outer,fillstyle=gradient,gradlines=2000,gradbegin=red,gradend=color717g,gradmidpoint=1.0](4.6521873,0.56671876){0.21}
\rput{-120.0}(11.912513,9.131128){\psframe[linewidth=0.01,dimen=outer,fillstyle=gradient,gradlines=2000,gradbegin=color2430g,gradend=color2430f,gradmidpoint=1.0](9.202186,1.1767182)(7.982186,1.0767182)}
\rput{120.0}(12.941419,-7.4182944){\psframe[linewidth=0.01,dimen=outer,fillstyle=gradient,gradlines=2000,gradbegin=color2430g,gradend=color2430f,gradmidpoint=1.0](9.222187,0.07671859)(8.002186,-0.023281408)}
\rput{-120.0}(10.182462,5.8922844){\psframe[linewidth=0.01,dimen=outer,fillstyle=gradient,gradlines=2000,gradbegin=color2430g,gradend=color2430f,gradmidpoint=1.0](7.402187,0.056718666)(6.1821866,-0.04328134)}
\psframe[linewidth=0.01,dimen=outer,fillstyle=gradient,gradlines=2000,gradbegin=color2430g,gradend=color2430f,gradmidpoint=1.0](8.302187,0.61671877)(7.082187,0.5167187)
\pscircle[linewidth=0.012,dimen=outer,fillstyle=gradient,gradlines=2000,gradmidpoint=1.0](7.092188,0.56671876){0.21}
\pscircle[linewidth=0.012,dimen=outer,fillstyle=gradient,gradlines=2000,gradbegin=red,gradend=color717g,gradmidpoint=1.0](8.272188,0.56671876){0.21}
\rput{120.0}(14.833997,-4.0709753){\psframe[linewidth=0.01,dimen=outer,fillstyle=gradient,gradlines=2000,gradbegin=color2430g,gradend=color2430f,gradmidpoint=1.0](9.202188,2.2967184)(7.9821877,2.1967185)}
\psframe[linewidth=0.01,dimen=outer,fillstyle=gradient,gradlines=2000,gradbegin=color2430g,gradend=color2430f,gradmidpoint=1.0](8.282187,2.8367188)(7.062187,2.7367187)
\rput{120.0}(13.068061,-1.1714116){\psframe[linewidth=0.01,dimen=outer,fillstyle=gradient,gradlines=2000,gradbegin=color2430g,gradend=color2430f,gradmidpoint=1.0](7.372188,3.2367184)(6.372188,3.1367185)}
\pscircle[linewidth=0.012,dimen=outer,fillstyle=gradient,gradlines=2000,gradmidpoint=1.0](7.0721874,2.7867188){0.21}
\rput{-120.0}(9.983143,12.057208){\psframe[linewidth=0.01,dimen=outer,fillstyle=gradient,gradlines=2000,gradbegin=color2430g,gradend=color2430f,gradmidpoint=1.0](8.972188,3.196719)(7.972187,3.096719)}
\pscircle[linewidth=0.012,dimen=outer,fillstyle=gradient,gradlines=2000,gradbegin=red,gradend=color717g,gradmidpoint=1.0](8.252187,2.7867188){0.21}
\rput{-120.0}(2.769886,6.0526314){\psframe[linewidth=0.01,dimen=outer,fillstyle=gradient,gradlines=2000,gradbegin=color2430g,gradend=color2430f,gradmidpoint=1.0](3.7421873,2.2767186)(2.5221872,2.1767185)}
\psframe[linewidth=0.01,dimen=outer,fillstyle=gradient,gradlines=2000,gradbegin=color2430g,gradend=color2430f,gradmidpoint=1.0](4.682187,2.8367188)(3.4621875,2.7367187)
\rput{120.0}(7.63342,1.8862802){\psframe[linewidth=0.01,dimen=outer,fillstyle=gradient,gradlines=2000,gradbegin=color2430g,gradend=color2430f,gradmidpoint=1.0](3.7721877,3.1967185)(2.7721877,3.0967185)}
\pscircle[linewidth=0.012,dimen=outer,fillstyle=gradient,gradlines=2000,gradmidpoint=1.0](3.4721875,2.7867188){0.21}
\rput{-120.0}(4.5358224,8.952196){\psframe[linewidth=0.01,dimen=outer,fillstyle=gradient,gradlines=2000,gradbegin=color2430g,gradend=color2430f,gradmidpoint=1.0](5.3521876,3.216719)(4.3521876,3.116719)}
\pscircle[linewidth=0.012,dimen=outer,fillstyle=gradient,gradlines=2000,gradbegin=red,gradend=color717g,gradmidpoint=1.0](4.6521873,2.7867188){0.21}
\rput{120.0}(9.241471,-7.5221286){\psframe[linewidth=0.01,dimen=outer,fillstyle=gradient,gradlines=2000,gradbegin=color2430g,gradend=color2430f,gradmidpoint=1.0](7.4021873,-1.0432813)(6.182187,-1.1432811)}
\rput{-120.0}(8.34509,2.6314778){\psframe[linewidth=0.01,dimen=outer,fillstyle=gradient,gradlines=2000,gradbegin=color2430g,gradend=color2430f,gradmidpoint=1.0](5.542187,-1.0432811)(4.3221874,-1.1432811)}
\psframe[linewidth=0.01,dimen=outer,fillstyle=gradient,gradlines=2000,gradbegin=color2430g,gradend=color2430f,gradmidpoint=1.0](6.4821877,-0.46328127)(5.2621875,-0.56328124)
\pscircle[linewidth=0.012,dimen=outer,fillstyle=gradient,gradlines=2000,gradmidpoint=1.0](5.272187,-0.5132812){0.21}
\pscircle[linewidth=0.012,dimen=outer,fillstyle=gradient,gradlines=2000,gradbegin=red,gradend=color717g,gradmidpoint=1.0](6.452187,-0.5132812){0.21}
\rput{-120.0}(13.865091,5.818451){\psframe[linewidth=0.01,dimen=outer,fillstyle=gradient,gradlines=2000,gradbegin=color2430g,gradend=color2430f,gradmidpoint=1.0](9.222188,-1.0432816)(8.002188,-1.1432816)}
\rput{-120.0}(15.552463,8.992655){\psframe[linewidth=0.01,dimen=outer,fillstyle=gradient,gradlines=2000,gradbegin=color2430g,gradend=color2430f,gradmidpoint=1.0](10.982187,0.056717902)(9.762188,-0.04328209)}
\rput{120.0}(14.641472,-10.63982){\psframe[linewidth=0.01,dimen=outer,fillstyle=gradient,gradlines=2000,gradbegin=color2430g,gradend=color2430f,gradmidpoint=1.0](11.002188,-1.0432811)(9.782187,-1.1432811)}
\psframe[linewidth=0.01,dimen=outer,fillstyle=gradient,gradlines=2000,gradbegin=color2430g,gradend=color2430f,gradmidpoint=1.0](10.122188,-0.46328127)(8.902187,-0.56328124)
\pscircle[linewidth=0.012,dimen=outer,fillstyle=gradient,gradlines=2000,gradmidpoint=1.0](8.912188,-0.5132812){0.21}
\pscircle[linewidth=0.012,dimen=outer,fillstyle=gradient,gradlines=2000,gradbegin=red,gradend=color717g,gradmidpoint=1.0](10.092187,-0.5132812){0.21}
\rput{120.0}(3.7987921,-4.3397965){\psframe[linewidth=0.01,dimen=outer,fillstyle=gradient,gradlines=2000,gradbegin=color2430g,gradend=color2430f,gradmidpoint=1.0](3.7621875,-1.0232815)(2.5421875,-1.1232815)}
\rput{120.0}(2.0040998,-1.1436301){\psframe[linewidth=0.01,dimen=outer,fillstyle=gradient,gradlines=2000,gradbegin=color2430g,gradend=color2430f,gradmidpoint=1.0](1.9421874,0.05671873)(0.7221875,-0.04328127)}
\rput{-120.0}(2.893129,-0.3962136){\psframe[linewidth=0.01,dimen=outer,fillstyle=gradient,gradlines=2000,gradbegin=color2430g,gradend=color2430f,gradmidpoint=1.0](1.9421875,-0.9832812)(0.7221875,-1.0832812)}
\psframe[linewidth=0.01,dimen=outer,fillstyle=gradient,gradlines=2000,gradbegin=color2430g,gradend=color2430f,gradmidpoint=1.0](2.8821874,-0.42328125)(1.6621875,-0.5232812)
\pscircle[linewidth=0.012,dimen=outer,fillstyle=gradient,gradlines=2000,gradmidpoint=1.0](1.6721874,-0.47328126){0.21}
\pscircle[linewidth=0.012,dimen=outer,fillstyle=gradient,gradlines=2000,gradbegin=red,gradend=color717g,gradmidpoint=1.0](2.8521874,-0.47328126){0.21}
\rput{120.0}(16.521368,-7.2451797){\psframe[linewidth=0.01,dimen=outer,fillstyle=gradient,gradlines=2000,gradbegin=color2430g,gradend=color2430f,gradmidpoint=1.0](10.962188,1.1967186)(9.7421875,1.0967185)}
\rput{-120.0}(13.569888,12.288014){\psframe[linewidth=0.01,dimen=outer,fillstyle=gradient,gradlines=2000,gradbegin=color2430g,gradend=color2430f,gradmidpoint=1.0](10.942188,2.276718)(9.722188,2.176718)}
\psframe[linewidth=0.01,dimen=outer,fillstyle=gradient,gradlines=2000,gradbegin=color2430g,gradend=color2430f,gradmidpoint=1.0](10.082188,1.7367188)(8.862187,1.6367188)
\pscircle[linewidth=0.012,dimen=outer,fillstyle=gradient,gradlines=2000,gradmidpoint=1.0](8.872188,1.6867188){0.21}
\pscircle[linewidth=0.012,dimen=outer,fillstyle=gradient,gradlines=2000,gradbegin=red,gradend=color717g,gradmidpoint=1.0](10.052187,1.6867188){0.21}
\rput{120.0}(3.9739966,2.1990495){\psframe[linewidth=0.01,dimen=outer,fillstyle=gradient,gradlines=2000,gradbegin=color2430g,gradend=color2430f,gradmidpoint=1.0](1.9621874,2.2967186)(0.74218744,2.196719)}
\rput{-120.0}(1.0098346,2.7964659){\psframe[linewidth=0.01,dimen=outer,fillstyle=gradient,gradlines=2000,gradbegin=color2430g,gradend=color2430f,gradmidpoint=1.0](1.9221874,1.1567187)(0.70218754,1.0567188)}
\psframe[linewidth=0.01,dimen=outer,fillstyle=gradient,gradlines=2000,gradbegin=color2430g,gradend=color2430f,gradmidpoint=1.0](2.8621874,1.7367188)(1.6421875,1.6367188)
\pscircle[linewidth=0.012,dimen=outer,fillstyle=gradient,gradlines=2000,gradmidpoint=1.0](1.6521875,1.6867188){0.21}
\pscircle[linewidth=0.012,dimen=outer,fillstyle=gradient,gradlines=2000,gradbegin=red,gradend=color717g,gradmidpoint=1.0](2.8321874,1.6867188){0.21}
\rput{120.0}(10.971523,-10.760974){\psframe[linewidth=0.01,dimen=outer,fillstyle=gradient,gradlines=2000,gradbegin=color2430g,gradend=color2430f,gradmidpoint=1.0](9.202188,-2.1632812)(7.9821873,-2.263281)}
\rput{-120.0}(12.09236,2.5149643){\psframe[linewidth=0.01,dimen=outer,fillstyle=gradient,gradlines=2000,gradbegin=color2430g,gradend=color2430f,gradmidpoint=1.0](7.3821874,-2.1832814)(6.1621876,-2.2832813)}
\psframe[linewidth=0.01,dimen=outer,fillstyle=gradient,gradlines=2000,gradbegin=color2430g,gradend=color2430f,gradmidpoint=1.0](8.282187,-1.5832813)(7.062187,-1.6832812)
\pscircle[linewidth=0.012,dimen=outer,fillstyle=gradient,gradlines=2000,gradmidpoint=1.0](7.092188,-1.6532812){0.21}
\pscircle[linewidth=0.012,dimen=outer,fillstyle=gradient,gradlines=2000,gradbegin=red,gradend=color717g,gradmidpoint=1.0](8.272188,-1.6532812){0.21}
\rput{120.0}(5.451523,-7.574){\psframe[linewidth=0.01,dimen=outer,fillstyle=gradient,gradlines=2000,gradbegin=color2430g,gradend=color2430f,gradmidpoint=1.0](5.5221868,-2.163281)(4.302187,-2.2632809)}
\rput{-120.0}(6.6450396,-0.5900472){\psframe[linewidth=0.01,dimen=outer,fillstyle=gradient,gradlines=2000,gradbegin=color2430g,gradend=color2430f,gradmidpoint=1.0](3.762188,-2.1632812)(2.542188,-2.2632813)}
\psframe[linewidth=0.01,dimen=outer,fillstyle=gradient,gradlines=2000,gradbegin=color2430g,gradend=color2430f,gradmidpoint=1.0](4.662187,-1.6032813)(3.4421875,-1.7032813)
\pscircle[linewidth=0.012,dimen=outer,fillstyle=gradient,gradlines=2000,gradmidpoint=1.0](3.4521875,-1.6532812){0.21}
\pscircle[linewidth=0.012,dimen=outer,fillstyle=gradient,gradlines=2000,gradbegin=red,gradend=color717g,gradmidpoint=1.0](4.632188,-1.6532812){0.21}
\rput{-120.0}(17.565039,5.714617){\psframe[linewidth=0.01,dimen=outer,fillstyle=gradient,gradlines=2000,gradbegin=color2430g,gradend=color2430f,gradmidpoint=1.0](11.042187,-2.1632814)(9.822186,-2.2632816)}
\psframe[linewidth=0.01,dimen=outer,fillstyle=gradient,gradlines=2000,gradbegin=color2430g,gradend=color2430f,gradmidpoint=1.0](10.122188,-2.7032812)(8.902187,-2.8032813)
\pscircle[linewidth=0.012,dimen=outer,fillstyle=gradient,gradlines=2000,gradmidpoint=1.0](8.912188,-2.753281){0.21}
\pscircle[linewidth=0.012,dimen=outer,fillstyle=gradient,gradlines=2000,gradbegin=red,gradend=color717g,gradmidpoint=1.0](10.092187,-2.753281){0.21}
\psframe[linewidth=0.01,dimen=outer,fillstyle=gradient,gradlines=2000,gradbegin=color2430g,gradend=color2430f,gradmidpoint=1.0](6.442188,-2.7032812)(5.222187,-2.8032813)
\pscircle[linewidth=0.012,dimen=outer,fillstyle=gradient,gradlines=2000,gradmidpoint=1.0](5.2321877,-2.753281){0.21}
\pscircle[linewidth=0.012,dimen=outer,fillstyle=gradient,gradlines=2000,gradbegin=red,gradend=color717g,gradmidpoint=1.0](6.412187,-2.753281){0.21}
\rput{120.0}(0.14616422,-4.430951){\psframe[linewidth=0.01,dimen=outer,fillstyle=gradient,gradlines=2000,gradbegin=color2430g,gradend=color2430f,gradmidpoint=1.0](1.9621875,-2.1232815)(0.74218744,-2.2232814)}
\psframe[linewidth=0.01,dimen=outer,fillstyle=gradient,gradlines=2000,gradbegin=color2430g,gradend=color2430f,gradmidpoint=1.0](2.9021876,-2.7032812)(1.6821874,-2.8032813)
\pscircle[linewidth=0.012,dimen=outer,fillstyle=gradient,gradlines=2000,gradmidpoint=1.0](1.6921874,-2.753281){0.21}
\pscircle[linewidth=0.012,dimen=outer,fillstyle=gradient,gradlines=2000,gradbegin=red,gradend=color717g,gradmidpoint=1.0](2.8721874,-2.753281){0.21}
\psframe[linewidth=0.01,dimen=outer,fillstyle=gradient,gradlines=2000,gradbegin=color2430g,gradend=color2430f,gradmidpoint=1.0](11.592187,0.61671877)(10.792188,0.5167187)
\pscircle[linewidth=0.012,dimen=outer,fillstyle=gradient,gradlines=2000,gradmidpoint=1.0](10.652187,0.5867188){0.21}
\psframe[linewidth=0.01,dimen=outer,fillstyle=gradient,gradlines=2000,gradbegin=color2430g,gradend=color2430f,gradmidpoint=1.0](11.632187,-1.6232812)(10.832188,-1.7232813)
\pscircle[linewidth=0.012,dimen=outer,fillstyle=gradient,gradlines=2000,gradmidpoint=1.0](10.692187,-1.6732812){0.21}
\psframe[linewidth=0.01,dimen=outer,fillstyle=gradient,gradlines=2000,gradbegin=color2430g,gradend=color2430f,gradmidpoint=1.0](11.612187,2.8367188)(10.812188,2.7367187)
\rput{120.0}(18.420742,-4.301783){\psframe[linewidth=0.01,dimen=outer,fillstyle=gradient,gradlines=2000,gradbegin=color2430g,gradend=color2430f,gradmidpoint=1.0](10.9521885,3.2167184)(9.9521885,3.1167185)}
\pscircle[linewidth=0.012,dimen=outer,fillstyle=gradient,gradlines=2000,gradmidpoint=1.0](10.652187,2.7867188){0.21}
\rput{-120.0}(-0.752216,5.7791452){\psframe[linewidth=0.01,dimen=outer,fillstyle=gradient,gradlines=2000,gradbegin=color2430g,gradend=color2430f,gradmidpoint=1.0](1.7921875,3.1567187)(0.7921876,3.0567186)}
\psframe[linewidth=0.01,dimen=outer,fillstyle=gradient,gradlines=2000,gradbegin=color2430g,gradend=color2430f,gradmidpoint=1.0](1.0521874,2.7367187)(0.5721875,2.6367188)
\pscircle[linewidth=0.012,dimen=outer,fillstyle=gradient,gradlines=2000,gradbegin=red,gradend=color717g,gradmidpoint=1.0](1.0921875,2.7067187){0.21}
\psframe[linewidth=0.01,dimen=outer,fillstyle=gradient,gradlines=2000,gradbegin=color2430g,gradend=color2430f,gradmidpoint=1.0](1.0121875,0.5767187)(0.53218746,0.47671875)
\pscircle[linewidth=0.012,dimen=outer,fillstyle=gradient,gradlines=2000,gradbegin=red,gradend=color717g,gradmidpoint=1.0](0.99218756,0.5467188){0.21}
\psframe[linewidth=0.01,dimen=outer,fillstyle=gradient,gradlines=2000,gradbegin=color2430g,gradend=color2430f,gradmidpoint=1.0](1.0521874,-1.5832813)(0.5721875,-1.6832812)
\pscircle[linewidth=0.012,dimen=outer,fillstyle=gradient,gradlines=2000,gradbegin=red,gradend=color717g,gradmidpoint=1.0](1.0121875,-1.6132812){0.21}
\psline[linewidth=0.06cm,linecolor=white,arrowsize=0.093cm 2.0,arrowlength=1.4,arrowinset=0.1]{->}(5.272187,-0.520217)(4.582187,0.6767188)
\psline[linewidth=0.06cm,linecolor=white,arrowsize=0.093cm 2.0,arrowlength=1.4,arrowinset=0.1]{->}(5.256399,-0.52129394)(6.637976,-0.5222043)
\psline[linewidth=0.06cm,linecolor=white,arrowsize=0.093cm 2.0,arrowlength=1.4,arrowinset=0.1]{->}(5.252976,-0.5237364)(4.5613995,-1.7197617)
\usefont{T1}{ptm}{m}{n}
\rput(5.934219,-0.22828124){\Large $\boldsymbol{\delta}_1$}
\usefont{T1}{ptm}{m}{n}
\rput(5.2,-1.2682812){\Large $\boldsymbol{\delta}_3$}
\usefont{T1}{ptm}{m}{n}
\rput(5.2,0.23171876){\Large $\boldsymbol{\delta}_2$}
\psline[linewidth=0.03cm,arrowsize=0.153cm 2.0,arrowlength=1.4,arrowinset=0.4]{->}(8.342187,0.6567188)(10.122188,1.6367188)
\usefont{T1}{ptm}{m}{n}
\rput(9.35,0.96){\Large $\vp{a}_1$}
\psline[linewidth=0.03cm,arrowsize=0.153cm 2.0,arrowlength=1.4,arrowinset=0.4]{->}(8.322187,0.55671877)(10.142188,-0.56328124)
\usefont{T1}{ptm}{m}{n}
\rput(9.35,0.15171875){\Large $\vp{a}_2$}
\usefont{T1}{ptm}{m}{n}
\rput(7.2262497,0.9517187){\Large A}
\usefont{T1}{ptm}{m}{n}
\rput(8.055469,0.93171877){\Large B}
\rput{120.0}(7.242819,-10.568206){\psframe[linewidth=0.01,dimen=outer,fillstyle=gradient,gradlines=2000,gradbegin=color2430g,gradend=color2430f,gradmidpoint=1.0](6.9721875,-3.1432812)(6.3721876,-3.2432811)}
\rput{120.0}(12.732819,-13.737858){\psframe[linewidth=0.01,dimen=outer,fillstyle=gradient,gradlines=2000,gradbegin=color2430g,gradend=color2430f,gradmidpoint=1.0](10.632187,-3.1432807)(10.032187,-3.2432806)}
\rput{120.0}(1.8728184,-7.467835){\psframe[linewidth=0.01,dimen=outer,fillstyle=gradient,gradlines=2000,gradbegin=color2430g,gradend=color2430f,gradmidpoint=1.0](3.3921876,-3.1432815)(2.7921872,-3.2432814)}
\rput{-120.0}(15.773745,2.7204125){\psframe[linewidth=0.01,dimen=outer,fillstyle=gradient,gradlines=2000,gradbegin=color2430g,gradend=color2430f,gradmidpoint=1.0](8.972188,-3.1432815)(8.372188,-3.2432814)}
\rput{-120.0}(10.253744,-0.46656057){\psframe[linewidth=0.01,dimen=outer,fillstyle=gradient,gradlines=2000,gradbegin=color2430g,gradend=color2430f,gradmidpoint=1.0](5.292187,-3.1432812)(4.692188,-3.2432811)}
\rput{-120.0}(4.8964233,-3.5196111){\psframe[linewidth=0.01,dimen=outer,fillstyle=gradient,gradlines=2000,gradbegin=color2430g,gradend=color2430f,gradmidpoint=1.0](1.7321875,-3.1232812)(1.1321874,-3.2232811)}
\psframe[linewidth=0.002,linecolor=white,dimen=outer,fillstyle=solid](11.642656,3.6967187)(0.46265626,3.5167189)
\psframe[linewidth=0.002,linecolor=white,dimen=outer,fillstyle=solid](11.662656,-3.3232813)(0.48265624,-3.503281)
\rput{-90.0}(11.495938,11.709374){\psframe[linewidth=0.002,linecolor=white,dimen=outer,fillstyle=solid](15.152656,0.186718)(8.052656,0.026718006)}
\rput{-90.0}(0.4359375,0.649375){\psframe[linewidth=0.002,linecolor=white,dimen=outer,fillstyle=solid](4.092656,0.16671877)(-3.0073438,0.04671876)}
\psline[linewidth=0.027999999cm,arrowsize=0.153cm 2.0,arrowlength=1.4,arrowinset=0.69]{<-}(0.3421875,-2.5432813)(0.3421875,-3.5632813)
\psline[linewidth=0.027999999cm,arrowsize=0.153cm 2.0,arrowlength=1.4,arrowinset=0.69]{<-}(1.2721875,-3.4932811)(0.2521875,-3.4932811)
\usefont{T1}{ptm}{m}{n}
\rput(1.5142186,-3.5282812){\Large $x$}
\usefont{T1}{ptm}{m}{n}
\rput(0.28421876,-2.3482811){\Large $y$}
\end{pspicture} 
}

        \label{fig_graphene_a}}
    \end{subfigure}
    \hfill
    \begin{subfigure}[]{
        \centering
     
\scalebox{0.4} 
{
\begin{pspicture}(0,-3.5967185)(10.060937,3.5967185)
\definecolor{color4}{rgb}{0.0,0.2,1.0}
\psline[linewidth=0.03cm,arrowsize=0.253cm 2.0,arrowlength=1.4,arrowinset=0.17]{<-}(4.2,2.8242188)(4.2,-2.4957812)
\psline[linewidth=0.03cm,arrowsize=0.253cm 2.0,arrowlength=1.4,arrowinset=0.17]{<-}(8.62,-0.0157813)(0.0,-0.0157813)
\usefont{T1}{ptm}{m}{n}
\rput(9.135156,-0.0207813){\huge $k_x$}
\usefont{T1}{ptm}{m}{n}
\rput(4.125156,3.2546873){\huge $k_y$}
\psline[linewidth=0.068cm,linecolor=color4](5.82,0.9442187)(5.82,-0.9757813)
\psline[linewidth=0.068cm,linecolor=color4](5.851384,0.9242187)(4.188616,1.8842187)
\psline[linewidth=0.068cm,linecolor=color4](4.188616,-1.9357812)(5.851384,-0.9757813)
\psline[linewidth=0.068cm,linecolor=color4](2.56,0.9442187)(2.56,-0.9757813)
\psline[linewidth=0.068cm,linecolor=color4](4.211384,-1.9157814)(2.548616,-0.9557813)
\psline[linewidth=0.068cm,linecolor=color4](2.528616,0.9042187)(4.1913843,1.8642187)
\psline[linewidth=0.04cm,linecolor=red,arrowsize=0.113cm 2.0,arrowlength=1.4,arrowinset=0.4]{->}(4.23,-0.028024487)(5.93,2.916462)
\psline[linewidth=0.04cm,linecolor=red,arrowsize=0.113cm 2.0,arrowlength=1.4,arrowinset=0.4]{->}(4.19,-0.003538114)(5.89,-2.9480245)
\usefont{T1}{ptm}{m}{n}
\rput(3.9479685,0.21921869){\Large \color{color4}$\boldsymbol\Gamma$}
\psdots[dotsize=0.17](5.82,-0.0157813)
\psdots[dotsize=0.17](5.82,0.9442187)
\psdots[dotsize=0.17](5.82,-0.9757813)
\psdots[dotsize=0.17](4.22,-0.0357813)
\usefont{T1}{ptm}{m}{n}
\rput(6.1854687,0.3992187){\LARGE \color{color4}$\vp{M}$}
\usefont{T1}{ptm}{m}{n}
\rput(5.865156,3.2592187){\huge \color{red}$\vp{b}_1$}
\usefont{T1}{ptm}{m}{n}
\rput(5.865156,-3.200781){\huge \color{red}$\vp{b}_2$}
\usefont{T1}{ptm}{m}{n}
\rput(6.2254686,1.4192187){\LARGE \color{color4}$\vp{K}'$}
\usefont{T1}{ptm}{m}{n}
\rput(6.2154684,-1.5007813){\LARGE \color{color4}$\vp{K}$}
\end{pspicture} 
}

       \label{fig_graphene_b}}
    \end{subfigure}
    \hfill
    \begin{subfigure}[]{
        \centering
      
\scalebox{0.4} 
{
\begin{pspicture}(0,-4.2079687)(9.414687,4.2479687)
\definecolor{color3g}{rgb}{1.0,0.8,0.0}
\definecolor{color3f}{rgb}{1.0,1.0,0.6}
\definecolor{color3b}{rgb}{1.0,1.0,0.8}
\definecolor{color215}{rgb}{1.0,0.2,0.2}
\rput{-180.0}(8.59375,1.6040626){\pstriangle[linewidth=0.04,dimen=outer,fillstyle=gradient,gradlines=2000,gradbegin=color3g,gradend=color3f,gradmidpoint=1.0,fillcolor=color3b](4.296875,-0.86796874)(2.72,3.34)}
\psellipse[linewidth=0.04,dimen=outer,fillstyle=solid](4.296875,2.4520311)(1.32,0.22)
\pstriangle[linewidth=0.04,dimen=outer,fillstyle=gradient,gradlines=2000,gradbegin=color3g,gradend=color3f,gradmidpoint=1.0,fillcolor=color3b](4.296875,-4.107969)(2.72,3.34)
\psellipse[linewidth=0.04,dimen=outer,fillstyle=solid](4.296875,-4.0979686)(1.34,0.11)
\psline[linewidth=0.024cm,arrowsize=0.153cm 2.0,arrowlength=1.4,arrowinset=0.31]{->}(4.296875,-0.7679688)(4.296875,3.7320313)
\psline[linewidth=0.024cm,arrowsize=0.153cm 2.0,arrowlength=1.4,arrowinset=0.31]{->}(4.306875,-0.77796876)(8.806875,-0.77796876)
\psline[linewidth=0.024cm,arrowsize=0.153cm 2.0,arrowlength=1.4,arrowinset=0.31]{->}(4.3254323,-0.7529687)(0.42831784,-3.0029688)
\psline[linewidth=0.06cm,linecolor=blue,arrowsize=0.093cm 2.0,arrowlength=1.4,arrowinset=0.4]{->}(4.296875,-0.80796874)(6.276875,-1.6679688)
\psline[linewidth=0.05cm,linecolor=color215,arrowsize=0.113cm 2.0,arrowlength=1.4,arrowinset=0.31]{->}(6.196875,-1.6279688)(7.016875,-1.9879688)
\psline[linewidth=0.05cm,linecolor=color215,arrowsize=0.113cm 2.0,arrowlength=1.4,arrowinset=0.31]{->}(6.216875,-1.5879687)(6.996875,-1.2879688)
\usefont{T1}{ptm}{m}{n}
\rput(7.1,-2.2729688){\huge \color{red}$\A{\vp{k}}$}
\usefont{T1}{ptm}{m}{n}
\rput(7.23,-0.99296874){\huge \color{red}$\A{\varphi}_{\vp{k}}$}
\usefont{T1}{ptm}{m}{n}
\rput(0.49421874,-3.2529688){\huge $k_x$}
\usefont{T1}{ptm}{m}{n}
\rput(8.824219,-0.41296875){\huge $k_y$}
\usefont{T1}{ptm}{m}{n}
\rput(4.284219,4.0270314){\huge $E_{\vp{k}}$}
\psbezier[linewidth=0.016,linecolor=blue](3.916875,-0.9428577)(4.0743217,-1.1479689)(4.656875,-1.1075244)(4.656875,-0.8879688)
\usefont{T1}{ptm}{m}{n}
\rput(3.6,-1.4){\huge $\varphi_{\vp{k}}$}
\usefont{T1}{ptm}{m}{n}
\rput(5.55,-1.0129688){\huge \color{blue}$\vp{k}$}
\end{pspicture} 
}

        \label{fig_graphene_c}}
    \end{subfigure}
 \caption{ The graphene lattice (a) and  its reciprocal lattice (b) , The Dirac points $\vp{K}'\left(\frac{2\pi}{3a},\frac{2\pi}{3\sqrt{3}a}\right)$ and $\vp{K}\left(\frac{2\pi}{3a},-\frac{2\pi}{3\sqrt{3}a}\right)$ are shown in the figure .The schematic plot of dispersion relation around the Dirac point (c)}
\label{fig:graphene} 
\end{figure*}

The high symmetry crystallographic points are presented in Fig.~\ref{fig_graphene_b}. 
Throughout this paper the graphene monolayer (laying on the xy-plane) interacts with a plane wave illuminating graphene in the perpendicular direction. This assumption allows us to use the electric dipole approximation in which the effect of magnetic field is excluded. Actually this approximation is quite accurate for an ideal graphene sheet wherein electrons are strongly bounded and their off-plane dynamics is negligible. The electric field can have an arbitrary time variation containing different harmonics. The dynamical properties of the positively charged ions that constitute the host lattice of the crystal will be neglected in our formulations. The system Hamiltonian for a single graphene sheet interacting with a classical electromagnetic field within the single particle approximation is
\begin{equation} \label{eq:H}
\A{H}=\A{H}_0+\A{H}_I
\end{equation}
Where $\A{H}_0$ governs the dynamics of the electrons with mass $M$ in the presence of the periodic lattice potential $V(\vp{r})$ 
\begin{equation}\label{eq:H0}
\A{H}_0=\int d^3\vp{r}\A{\Psi}^\dagger(\vp{r})\left\{\frac{\A{\vp{p}}^2}{2M}+V(\vp{r})\right\}\A{\Psi}(\vp{r})
\end{equation}
The interaction Hamiltonian in long-wavelength limit (or normal incidence) is rigorously obtained in the \textit{velocity-gauge} by replacing $\vp{p}$ by $\vp{p}+e\vp{A}$ in $\A{H}_0$ where $\vp{A}$ is the associated vector magnetic potential \cite{aversa1995nonlinear}. In the case of graphene it seems very simple to use this electrodynamics substitution, however it can be shown that neither the calculations are efficient, nor it reveals some interesting physical properties. The interaction problem can be recast into the \textit{length gauge}
\begin{equation}
\A{H}_{I}=e\vp{E}_{in}(t)\cdot\int d^3\vp{r}\A{\Psi}^\dagger(\vp{r}) \vp{r}\A{\Psi}(\vp{r})
\end{equation}
A common problem with the perturbation theory for solids in the length gauge is the difficult treatment of the position operator $\vp{r}$ in view of the extended Bloch states \cite{aversa1995nonlinear}. This is further detailed in the next section. It will be shown that both $\A{H}_0$ and $\A{H}_I$ can be treated in the framework of the Tight-Binding (TB) regime. Following the TB approximation the field operator is expanded in terms of $2p_z$ orbital wavefunction $\phi(\vp{r})$
\begin{eqnarray}\label{eq:TB-Field Operator}
\A{\Psi}(\vp{r}) &=& \frac{1}{\sqrt{N}}\sum_{\vp{k}}\sum_{\vp{R}_A}\exp(i\vp{k}\cdot\vp{R}_A)\phi(\vp{r}-\vp{R}_A)\A{a}_{\vp{k}}\nonumber\\
&&\nonumber \\
&&
+\frac{1}{\sqrt{N}}\sum_{\vp{k}}\sum_{\vp{R}_B}\exp(i\vp{k}.\vp{R}_B)\phi(\vp{r}-\vp{R}_B)\A{b}_{\vp{k}}\nonumber\\
&&\nonumber\\
&=& \A{\Psi}_A(\vp{r}) +\A{\Psi}_B(\vp{r}) 
\end{eqnarray}
The summation run over all sublattices coordinates denoted by $\vp{R}_{A/B}$. The operators $\A{a}_{\vp{k}}$ and $\A{b}_{\vp{k}}$ are fermionic annihilation operators on A and B sublattices respectively. The TB hopping parameters can be found by fitting the results of the first-principles electronic structure calculations with the experimental results. The simplest TB Hamiltonian contains hopping to the nearest-neighbor sites. The resulted Hamiltonian is easy to diagonalized in the second quantized operators $\A{a}_{\vp{k}}$ and $\A{b}_{\vp{k}}$. Inserting \eqref{eq:TB-Field Operator} into \eqref{eq:H0} and taking into account the nearest-neighbor hopping $\kappa\approx -2.97\mathrm{eV}$ \cite{reich2002tight} along $\bo{\delta}_1$, $\bo{\delta}_2$ and $\bo{\delta}_3$ bonds (shown in Fig.\ref{fig:graphene}) the TB Hamiltonian in the momentum space is
\begin{multline} 
\label{eq:TB-H0}
\A{H}_0=\sum_{\vp{k}}E_{0}\left(\A{a}_{\vp{k}}^\dagger\A{a}_{\vp{k}}+\A{b}_{\vp{k}}^\dagger\A{b}_{\vp{k}}\right)\\
+\kappa\sum_{\vp{k}}\left[f(\vp{k})\A{a}^\dagger_{\vp{k}}\A{b}_{\vp{k}}+f^*(\vp{k})\A{b}_{\vp{k}}^\dagger{a}_{\vp{k}}\right]
\end{multline} 
$E_0$ is the result of the hopping processes within the sublattices. The first term on the right hand side of the Eq.~ \eqref{eq:TB-H0} is symmetrically diagonal and does not affect the quasi-particles dynamics. The function $f(\vp{k})$ carries the symmetry properties of the graphene lattice and is given by 
\begin{equation}
f(\vp{k})=\sum_{i=1}^{3}\exp(i\vp{k}.\bo{\delta}_i)
\end{equation}
The TB Hamiltonian $\A{H}_0$ becomes diagonal in the conduction and valence basis
\begin{align}
&\A{\xi}_{\vp{k}c}=\frac{1}{\sqrt{2}}\left(e^{-i\alpha_{\vp{k}}/2}\A{a}_{\vp{k}}+e^{+i\alpha_{\vp{k}}/2}\A{b}_{\vp{k}}\right)\\
&\A{\xi}_{\vp{k}v}=\frac{1}{\sqrt{2}}\left(e^{-i\alpha_{\vp{k}}/2}\A{a}_{\vp{k}}-e^{+i\alpha_{\vp{k}}/2}\A{b}_{\vp{k}}\right)
\end{align} 
where $f(\vp{k})=\abs{f(\vp{k})}\exp(i\alpha_{\vp{k}})$. This yields the corresponding TB-based band structure
\begin{equation}
E^{c/v}_{\vp{k}}=E_0\pm\kappa \abs{f(\vp{k})}
\end{equation} 
It is easy to show that the two bands cross at $\vp{K}$ and $\vp{K}'$ and the Hamiltonian displays the interesting physics of the Dirac fermions in the proximity of the conical points. These special crossing points are topologically protected and no band gap can be opened in the presence of any kind of perturbation preserving time reversal and inversion symmetries. For the case of intrinsic graphene sheet the Fermi surface coincides with the energy at the conical points. For a reasonably doped graphene sheet, just the dynamics of the quasiparticles around the conical points play the major role in the electronic properties of graphene. In this scenario, graphene can be viewed as a vanishing-gap semiconductor. Some other crystalline  structures such as $\mathrm{HgTe}$ are also known to be gapless semiconductors \cite{Tsidilkovski}, but what makes graphene unique is also the helical nature of the quasiparticles \cite{katsnelson}. In the next sections it will be shown that this helicity plays a decisive role in nonlinear optical response of graphene which is the focus of this paper. 
In the vicinity of the conical points $f(\vp{k})$ can be linearly expanded in terms of $\vp{k}$ components as 
\begin{equation}
f(\vp{K}+\vp{k})\approx -\frac{3}{2}ka e^{-i(\pi/6\pm\varphi_{\vp{k}})}
\end{equation}
where $\varphi_{\vp{k}}$ is the angle of vector $\vp{k}$ with respect to the $k_x$ axis shown in Fig.~\ref{fig_graphene_c}. Linearization of the Hamiltonian around the conical points yields a massless Dirac quasiparticle whose dispersion relation is $E_{\vp{k}}=\pm \hbar v_F k$. It is noted that $v_F=- 3a\kappa/2\hbar$ is the Fermi velocity which is around $c/300$. Within the band structure picture the effective Hamiltonian is 
\begin{equation}
\A{\mathcal{H}}\approx \sum_{\vp{k}}\hbar v_F k\left(\A{\xi}_{\vp{k}c}^\dagger\A{\xi}_{\vp{k}c}-\A{\xi}_{\vp{k}v}^\dagger\A{\xi}_{\vp{k}v}\right)
\end{equation}
where $\A{\xi}_{\vp{k}c}$ and $\A{\xi}_{\vp{k}v}$ are the conduction and valence annihilation operators in the upper and lower energy bands, respectively. As long as the inter-valley scattering is improbable the local behavior of the Hamiltonian around $\vp{K}$ and $\vp{K}'$ is independent. For mathematical convenience the matrix representations are used in the derivation of the equations of motion. In fact, the system can be adequately described in \textit{atomistic} language. Assume that two-component spinors
$\begin{pmatrix} 
1 & 0
\end{pmatrix}^T
$ 
and 
$\begin{pmatrix} 
0 & 1
\end{pmatrix}^T
$ 
are adopted for $A$ and $B$ states respectively. The resulting time dependeted Dirac equation describing low energy excitation around one of the conical points is written as 
\begin{align}
&\A{\mathcal{H}}_{\vp{k}}=\hbar v_F\vp{k}\cdot\vec{\sigma}\nonumber \\
&i\hbar\pds{t}\overline{\Psi}_{\vp{k}}=\A{\mathcal{H}}_{\vp{k}}\overline{\Psi}_{\vp{k}}\label{eq:314}
\end{align}
where $\overline{\Psi}_{\vp{k}}$ is a two-component spinor in the momentum space and $\vec{\sigma}$ is made from the Pauli matrices $\vec{\sigma}=\A{x}\sigma_x+\A{y}\sigma_y+\A{z}\sigma_z$ arose form the two sublattices. 
The optical response of graphene can be studied in a physically transparent way using the equation of motion for the density matrix. In Sec.~\ref{section:4} we examine the nonlinear optical response of graphene based on the time evolution of the density matrix. Liouville's equation governs the time progress of the density matrix  and it yields coupled differential equations.
\begin{equation}\label{eq:Liouville}
i\hbar \pds{t}\A{\rho}_{\vp{k}}=[\A{H},\A{\rho}_{\vp{k}}]=[\A{\mathcal{H}}_{\vp{k}},\A{\rho}_{\vp{k}}]+[\A{H}_I,\A{\rho}_{\vp{k}}]
\end{equation} 
The first term on the right hand side of Eq.~\eqref{eq:Liouville} is  the regular dynamical phase variation. In the next section it will be shown that the second term is closely related to the Berry connection and topological properties of the band structure. This is the point where chiral nature of the carriers and strong optical response of graphene tie up. In the next section, an intuitive clue will be provided in a general formalism of two-band systems.

\section{Two-Band Systems and The Role of Chirality\label{section:3}}
To illustrate the impact of the chiral nature of the charged carriers on the optical response of graphene and to explore the uniqueness of the graphene in terms of its strong nonlinear interaction with light, the mathematical description of chirality for a general two-level systems is presented. We also address applicability of the reduced TB basis to describe the matrix elements of the interaction Hamiltonian in the length gauge and its connection with chirality of the charged carriers.
In the last part of this section, it is shown that, our arguments are general enough and they are independent of the approximation existing in TB calculations. Chirality and its influence on the optical response root in the discrete symmetries existing in the crystalline structure of graphene and inclusion of the many body effects and the other higher order interaction terms do not alter the general conclusion.

\subsection{Two-band systems}
A prototypical two-band system might be described by the Hamiltonian expanded in terms of the Pauli matrices as
\begin{equation}
\A{H}=\varepsilon_0(\vp{k})\A{I}+\varepsilon(\vp{k})\vec{u}(\vp{k}).\A{\vec{\sigma}}
\end{equation}
where $\varepsilon_0(\vp{k})$ and $\varepsilon(\vp{k})>0$ are real functions of Bloch wavenumber $\vp{k}$. The three-dimensional vector operator $\A{\vec{\sigma}}$ is made from the Pauli matrices and $\A{I}$ is a $2\times 2$ identity matrix. The vector $\vec{u}(\vp{k})$ is a three dimensional unit vector and can be represented in terms of the spherical angle variables $\alpha$ and $\beta$
\begin{align}
&u_x(\vp{k})=\sin\beta(\vp{k})\cos\alpha(\vp{k})\label{eq:ux}\\
&u_y(\vp{k})=\sin\beta(\vp{k})\sin\alpha(\vp{k})\label{eq:uy}\\
&u_z(\vp{k})=\cos\beta(\vp{k})\label{eq:uz}
\end{align} 
By successive applications of the rotation operator $\mathcal{D}(\A{\vp{n}},\phi)=\exp(-\frac{1}{2}\A{\vp{n}}.\vec{\sigma}\phi)$ ($\A{\vp{n}}$ and $\phi$ are the axis and the angle of the rotation respectively) along the Euler axis \cite{sakurai} the eigenvectors correspond to two energy eigenvalues 
$E_{\pm}(\vp{k})=\varepsilon_0(\vp{k})\pm\varepsilon(\vp{k})$
of the Hamiltonian can be obtained 
\begin{align}
&\ket{\vp{k},\uparrow}=
\begin{bmatrix}
\cos\left(\frac{\beta}{2}\right)e^{-i\alpha/2}\\
\sin\left(\frac{\beta}{2}\right)e^{+i\alpha/2}
\end{bmatrix}
\label{eq:alpha-beta1}\\
&\ket{\vp{k},\downarrow}=
\begin{bmatrix}
\sin\left(\frac{\beta}{2}\right)e^{-i\alpha/2}\\
-\cos\left(\frac{\beta}{2}\right)e^{+i\alpha/2}
\label{eq:alpha-beta2}
\end{bmatrix}
\end{align}
For a $d$-dimensional electronic system the Bloch momentum $\vp{k}$ can be represented by its magnitude $k$ and $d-1$ angle variables in the spherical coordinates, $\left\{\gamma_1,\gamma_2,\cdots,\gamma_{d-1}\right\}$. For the sake of brevity all angle variables are conveniently called $\gamma$. For the particular case of graphene the system is two-dimensional and only the azimuthal angle $\varphi_{\vp{k}}$ is needed to determine the direction of the Bloch momentum in the reciprocal space. Suppose a prototype Hamiltonian in which $\vec{u}$ 
is a function of angle variables only, i.e. $\vec{u}=\vec{u}(\gamma)$, with no dependency on $k$, this is known as a \textit{general chiral system} \cite{bacsi2013}. For such a system $\alpha$ and $\beta$ appearing  in Eqs.~ \eqref{eq:alpha-beta1} and \eqref{eq:alpha-beta2} only depend on the angle variables. Equivalently  the pseudospin is determined by the direction of the momentum. 

According to Eq.~\eqref{eq:314} it is obvious that the low energy Hamiltonian in graphene describes a scale invariance chiral electronic system. The chiral symmetry of the carriers is not restricted to the TB model but stems from the honeycomb translational symmetry of the crystalline structure.

\subsection{Position operator: The role of chirality}
To examine the importance of chirality in the optical response of graphene, we  now turn our attention to the calculation of matrix elements of the interaction Hamiltonian in the length gauge. As mentioned earlier, the calculation of matrix elements of the position operator in different Bloch states is challenging and it has caused some controversies \cite{aversa1995nonlinear,gradhand2012first}. For the general case of extended Bloch states with spatial dependency of $\Psi_{n\vp{k}}(\vp{r})=\bra{\vp{r}}n,\vp{k}\rangle=\exp(i\vp{k}.\vp{r})u_{n\vp{k}}(\vp{r})$ where $u_{n\vp{k}}$ is the periodic part of the wavefunction, the matrix elements of the position operator are related to Berry connection tensor \cite{gradhand2012first}. It is also shown \cite{gradhand2012first} that $[\A{\vp{r}},\A{\rho}_{\vp{k}}]$ appeared on the left hand side of Eq.~\eqref{eq:Liouville} can be expressed as
\begin{equation}\label{eq:160}
[\A{\vp{r}},\A{\rho}_{\vp{k}}]=-i\nabla_{\vp{k}}\A{\rho}_{\vp{k}}+[\OO{\vp{A}}_{\vp{k}},\A{\rho}_{\vp{k}}]
\end{equation}
where $\OO{\vp{A}}_{\vp{k},nm}=-i\bra{u_{n,\vp{k}}}\nabla_{\vp{k}}\ket{u_{m,\vp{k}}}$ is the Berry connection tensor. Actually, the most difficulty in calculation of the matrix elements of the position operator is due to difficulty in the calculations of Berry connection. In order to compute the Berry connection tensor rigorously, the full machinery of Density Function Theory and Wannier interpolation scheme should be used \cite{gradhand2012first}. Introducing maximally localized Wannier basis functions provides a numerically feasible scheme to evaluate the matrix elements. It is straightforward to show that in the proper gauge that the basis functions are expanded around the atomic centers $\vp{r}_\alpha$ as 
\begin{equation*}
\Psi_{n,\vp{k}}(\vp{r})=\sum_{\vp{R},\alpha}C_{\alpha,n\vp{k}}\exp\left[i\vp{k}.(\vp{R}+\vp{r}_{\alpha})\right]\phi_{\alpha}(\vp{r}-\vp{R}-\vp{r}_{\alpha})
\end{equation*}
the right hand side of Eq.~\eqref{eq:160} reads
\begin{equation}\label{eq:1011}
[\A{\vp{r}},\A{\rho}_{\vp{k}}]=-i\nabla_{\vp{k}}\A{\rho}_{\vp{k}}+[\OO{\boldsymbol{\zeta}}_{\vp{k}},\A{\rho}_{\vp{k}}]
\end{equation}
where 
$\boldsymbol{\zeta}_{\vp{k},nm}=-i\sum_{\alpha}C_{\alpha,n\vp{k}}\nabla_{\vp{k}}C_{\alpha,m\vp{k}}$ is closely related to the Berry connection tensor. 
Localization of basis functions is the basic principle that underlines this approximation. Fortunately, for the case of graphene the basis functions are fairly well localized and this approximation works well. It is worth mentioning that both terms appearing on the right hand side of Eq.~\eqref{eq:1011} are gauge dependent, but the overall expression is independent of gauge and the specific choice of basis functions.
For the particular case of two band chiral systems described in the previous section, the Berry connection exhibits singular behavior at the degeneracy points. Energy eigenstates only depend on the angular variables $\gamma_i$'s. Therefore the gradient operator acting on the angular functions will be 
\begin{equation}
\nabla_{\vp{k}}C_{\alpha,\vp{k}n}=\frac{1}{k}\sum_{i}\A{\gamma}_i\frac{1}{h_i(\gamma)}\pds{\gamma_i}C_{\alpha,\vp{k}n}
\end{equation} 
where $kh_i(\gamma)$ and $\A{\gamma}_i$ are respectively, the Riemann scale function and the unit vector associated with the angle variable $\gamma_i$. The most interesting property of the chiral systems in their optical response originate from the $1/k$ dependence. The appearance of $1/k$ term in the Liouville equation is the main difference between graphene and an ordinary semiconductor material. 
It is shown in the next section that this term acts like a dipole in the reciprocal space, playing a significant role in the graphene's nonlinear optical response.

\subsection{Chirality and symmetry\label{subsection:C}}
 Basically, low energy excitations that capture the universal characteristics of the system are highly influenced by symmetries. Let us focus on the effective Hamiltonian governing electrons dynamics around the Fermi energy level including band renormalizations due to electron-electron interactions. In this paper we will not plunge into the Landau theory and just symmetry considerations are discussed here. As long as the two Dirac points can be treated as independent entities, first order expansion of the Hamiltonian around the conical points reads
\begin{equation}
\A{H}^{eff}_{\vp{k}}\approx \sum_{i,j}k_i A_{ij}\A{\sigma}_j+ m(\vp{k})\A{\sigma}_z
\end{equation}
where $i$ and $j$ run over $x$ and $y$. The coefficients $A_{ij}$'s are the elements of a $2\times 2$ constant real matrix. The mass term $m(\vp{k})$ can be expanded as $m(\vp{k})=m_0+k_xm_x+k_ym_y$. Based on the mathematical description of the chirality elucidated above, it is easy to show that the necessary and sufficient condition to have a chiral system in low energy limit is $m_0=0$, which implies  gapless state. It can be shown that the Dirac fermions (gapless property) are topologically protected \cite{bernevig,hatsugai2011topological}.
Two main symmetries characterize the hexagonal lattice with identical atoms on A and B sites. The first one is time reversal which exists in the absence of the magnetic interactions and complex hopping. This symmetry is present regardless of space symmetry properties of atomic potential. The second symmetry, inversion, is induced by the mirror symmetry of the atomic arrangement. Clearly this symmetry is present in the isotropic hopping scenario.
It is easy to show that time reversal and inversion symmetries separately establish a relationship between the Hamiltonians for different valleys and 
only if both symmetries are preserved $m_0$ must necessarily vanish and the Dirac nodes would be locally stable.
Any kind of perturbation satisfying these symmetries cannot open a gap as long as the Dirac points do not meet up. Moreover, two inequivalent Dirac points carry vortices with opposite signs and the nodes with opposite vortices cannot be removed by themselves. For large enough perturbation, however, two different Dirac points may meet each other and annihilate the vortices and open up a gap. It can be shown that in the presence of $C_{3h}$ symmetry the gapless Dirac points are globally stable \cite{bernevig}. Taking all these symmetries into account the low energy Hamiltonian describes chiral quasiparticles around the Dirac points.

In the subsequent sections it will be shown that this chiral behavior leads to a singularity in higher order optical response of graphene. To remove this singular behavior while keeping the symmetries intact a suitable band renormalizations can be used. In 2005, Kane and Mele showed that at sufficiently low energy an isolated graphene exhibits a quantum spin Hall effect with an energy gap induced by spin-orbit interaction \cite{kane2005quantum}. Although this band gap is very small, graphene as a critical electronic state can be strongly affected by this perturbation. Taking all contributions into account, the spin-orbit coupling Hamitonian is
\begin{equation}
\A{H}_{so}=-\Delta_{so}\A{\sigma}_z\A{\tau}_z\A{s}_z+\lambda_R\left( \A{\sigma}_x\A{\tau}_z\A{s}_y- \A{\sigma}_y\A{\tau}_0\A{s}_x\right)
\end{equation}
where $\A{\sigma}_i$ , $\A{\tau}_i$ and $\A{s}_i$ are Pauli matrices acting on pseudospin, valley index and electron spin respectively. The coefficients $\Delta_{so}$ and $\lambda_R$ are microscopic spin-orbit coupling constant and Rashba coefficient, (as a result of breaking mirror symmetry) respectively. The spin-orbit coupling factor $\Delta_{so}$ can be affected by curvature of the graphene sheet. The reported value for this coefficient for the ideal case of flat defect-free graphene is $\Delta_{so}\approx 1 \mathrm{\mu eV}$ \cite{choudhari2014graphene}. For all practical structures $\lambda_R$ is much smaller than $\Delta_{so}$ and the  resulting energy gap is $2(\Delta_{so}-\lambda_R)$. This small gap can remove the singularity in the optical response of graphene which will be discussed in the next sections.

\section{Semiconductor Bloch Equations for graphene\label{section:4}}
As mentioned in the preceding sections, within the single particle approximation, the density matrix obeys dynamical equations in Schr\"odinger's picture. The applied  field drives the distribution out of equilibrium leading to nonvanishing induced current.
The dynamical equations on the density matrix can be recast into the form of the Semiconductor Bloch Equations (SBEs). 
Based on SBEs the dynamics is governed by a quasiclassical theory with quantum fluctuations superimposed. The quantum corrections to the classical dynamics will be converted to the well known problem of light-atom interaction \cite{boyd2003nonlinear}.

\subsection{Equations of motion}
We proceed from Eq.~\eqref{eq:160} which offers a gauge independent relation and thus we are at liberty to choose any kind of gauge making the mathematical structure simpler. Working in the sublattice (pseudospin) basis and making use of Eq.~\eqref{eq:Liouville} and Eq.~\eqref{eq:1011} gives
\begin{equation}\label{eq:2000}
i\hbar\pd{\A{\rho}_{\vp{k}}}{t}=\hbar v_F \vp{k}.[\A{\vec{\sigma}},\A{\rho}_{\vp{k}}]+ie\vp{E}.\nabla_{\vp{k}}\A{\rho}_{\vp{k}}
\end{equation}
Due to the smallness of the band gap induced by spin-orbit coupling, the dispersion properties of the charged carriers would barely deviate from massless relativistic dynamics and can be safely neglected around the Dirac points. However, we have to acknowledge the importance of this effect in the nonlinear optical response of graphene. The phenomenological inclusion of spin-orbit coupling as well as scattering due to imperfections will be discussed at the end of this section. 
The $2\times 2$ pseudospin density matrix $\A{\rho}_{\vp{k}}$ can be expanded in terms of Pauli matrices
\begin{equation}\label{eq:101}
\A{\rho}_{\vp{k}}=n_{\vp{k}}\A{I}+\vec{m}_{\vp{k}}\cdot\A{\vec{\sigma}}
\end{equation}
On substituting Eq.~\eqref{eq:101} into Eq.~\eqref{eq:2000}, one obtains decoupled equations for \textit{charge density} $n_{\vp{k}}$ and \textit{pseudospin density} $\vec{m}_{\vp{k}}$ \cite{katsnelson}
\begin{align}
&\pd{n_{\vp{k}}}{t}=\frac{e}{\hbar}\vp{E}.\nabla_{\vp{k}} n_{\vp{k}}\label{eq:110}\\
&\pd{\vec{m}_{\vp{k}}}{t}=2v_F\left(\vp{k}\times \vec{m}_{\vp{k}}\right)+\frac{e}{\hbar}\vp{E}.\nabla_{\vp{k}}\vec{m}_{\vp{k}}\label{eq:111}
\end{align} 
The right hand side of Eq.~\eqref{eq:111} is analogous to spin procession in a magnetic field. The same can be set for the pseudospin in the \textit{psudomagnetic field} acting in the reciprocal space \cite{katsnelson}. This equation encodes a wealth of information about the optical response of graphene including linear and nonlinear response in noninteracting regime. Owing to the linear dispersion relation around the Dirac points, current operator has only paramagnetic component
\begin{equation}
\A{\vec{\mathcal {J}}}_{\vp{k}} =-\frac{e}{\hbar}\pd{\A{\mathcal{H}}_{\vp{k}}}{\vp{k}}=-ev_F\A{\vec{\sigma}}
\end{equation}
and the current density becomes
\begin{equation}
\vp{J}=\Ex{\A{\vec{\mathcal {J}}}_{\vp{k}}}=\Tr\left(\A{\vec{\mathcal {J}}}_{\vp{k}}\A{\rho}_{\vp{k}}\right)=-2ev_F\left(\A{x}\A{x}+\A{y}\A{y}\right)\cdot\sum_{\vp{k}}\vec{m}_{\vp{k}}
\end{equation}
Having derived the equations of motion in the sublattice basis, now, we can switch to the energy diagonal basis. To avoid confusion, we use ``$\sim$'' to denote the matrix representation of the operators in the valence and conduction basis. In the energy diagonal basis
$
\begin{pmatrix}
1 &0
\end{pmatrix}^T
$
and 
$
\begin{pmatrix}
0 &1 
\end{pmatrix}^T
$,
stand for the upper and the lower energy levels respectively. In the energy diagonal basis the density matrix and the current operator become:
\begin{align}
\tilde{\rho}_{\vp{k}}=&\A{I}n_{\vp{k}}+\vec{m}_{\vp{k}}.\left(\A{\vp{k}}\sigma_z +\A{\varphi}_{\vp{k}}\sigma_{y}-\A{\vp{z}}\sigma_x \right)\label{eq:rho100}\\
\tilde{\vec{\mathcal{J}}}_{\vp{k}}=&-ev_F\left(\A{\vp{k}}\sigma_z+\A{\varphi}_{\vp{k}}\sigma_y\right)\label{eq:J100}
\end{align}
Where $\A{\vp{k}}$ and $\A{\varphi}_{\vp{k}}$ are shown in Fig.~\ref{fig_graphene_c}. In the thermal equilibrium, before switching on the incident field, the density distribution obeys fermion statistics
\begin{equation}
\Ex{\A{\xi}_{\vp{k}c}^\dagger\A{\xi}_{\vp{k}c}}_0 = f(\mathcal{E}(\vp{k})) \quad,\quad\Ex{\A{\xi}_{\vp{k}v}^\dagger\A{\xi}_{\vp{k}v}}_0 = f(-\mathcal{E}(\vp{k})) 
\end{equation} 
Where subscript $0$ denotes equilibrium state and $\mathcal{E}(\vp{k})=\hbar v_f k$ is the upper energy level. The distribution $f(E)$ is the fermionic distribution function
\begin{equation*}
f(E)=\frac{1}{1+\exp\left(\frac{E-\mu}{K_BT}\right)}
\end{equation*}
where $\mu$ and $T$ are, respectively, the chemical potential associated with the fermi energy level $E_f$ and 
the temperature. 
\subsection{Dynamics of population difference and polarization}
In the presence of electromagnetic field, the current operator acquires a finite expectation value. The particle current, can be divided into two distinct parts. The first part is the current resulting from disturbing the distribution of the charged carriers residing on the upper and lower energy levels and the second contribution is due to interference between them. The former is intraband and the latter is the interband current. Following this statement, it will be shown that the optical response of a general two level system depends on the population difference $\mathcal{N}(\vp{k})$ and polarization $\mathcal{P}(\vp{k})$
\begin{align}
&\mathcal{N}(\vp{k},t)=\Ex{\A{\xi}_{\vp{k}c}^\dagger\A{\xi}_{\vp{k}c}}-\Ex{\A{\xi}_{\vp{k}v}^\dagger\A{\xi}_{\vp{k}v}}=2\A{\vp{k}}.\vec{m}\label{def:N}\\
&\mathcal{P}(\vp{k},t)=\Ex{\A{\xi}_{\vp{k}v}^\dagger\A{\xi}_{\vp{k}c}}=-\A{\vp{z}}.\vec{m}+i\A{\varphi}_{\vp{k}}.\vec{m}\label{def:P}
\end{align}
Taking $\mathcal{N}$ and $\mathcal{P}$ as dynamical varibales and using Eq.~\eqref{eq:111}, we obtain the equations of motion for the population difference and the polarization.
\begin{widetext}
\begin{equation}\label{GSBE}
\left\{
\begin{array}{l}
\pd{\mathcal{N}(\vp{k},t)}{t}-\frac{e}{\hbar}\vp{E}.\nabla_{\vp{k}}\mathcal{N}(\vp{k},t)=-2\Phi(\vp{k},t)\mathrm{Im}\left\{\mathcal{P}(\vp{k},t)\right\}\\
\\
\pd{\mathcal{P}(\vp{k},t)}{t}-\frac{e}{\hbar}\vp{E}.\nabla_\vp{k}\mathcal{P}(\vp{k},t)=i\Omega(\vp{k})\mathcal{P}(\vp{k},t)
+\frac{i}{2}\Phi(\vp{k},t)\mathcal{N}(\vp{k},t)
\end{array}
\right.
\end{equation}
\end{widetext}
where $\Phi(\vp{k},t)$ is the effective dipole causing interband transitions in the reciprocal space and $\Omega(\vp{k})$ is the Rabi frequency associated with interband transitions 
\begin{align}
& \Phi(\vp{k},t)=\frac{e}{\hbar}\frac{\vp{E}\cdot\A{\varphi}_{\vp{k}}}{k}\\
&\Omega(\vp{k})=\frac{2\mathcal{E}(\vp{k})}{\hbar}=2v_Fk
\end{align}
The coupled equations given in \eqref{GSBE} are called semiconductor Bloch equations (SBEs) for graphene. These equations must be solved simultaneously, subject to the initial condition imposed by the fermion distribution before turning on the field. 
\begin{align*}
&\mathcal{N}(\vp{k},-\infty)=f(\mathcal{E}(\vp{k}))-f(-\mathcal{E}(\vp{k}))\\
&\mathcal{P}(\vp{k},-\infty)=0
\end{align*}
The left side of the SBEs are essentially similar to the semiclassical Boltzman's transport equation. This part of dynamics is responsible for intraband transitions for a pure graphene, neglecting the effect of collisions and imperfections. A simple way of incorporating the effect of collision into the theory is to use a complex frequency in the spectral domain. The right side of SBEs appear to resemble the problem of two level atomic transition in the presence of an applied electric field. However, the dipole that causes transition has been replaced by $\Phi(\vp{k},t)$ in the reciprocal space. The chiral nature of the charged carriers leaves its fingerprint on the appearance of $1/k$ in the effective dipole expression. This singular behavior roots in $1/k$ dependence of $\OO{\bs{\zeta}}_{\vp{k}}$ in the energy diagonal basis. As discussed earlier, this singularity can be resolved by the phenomenological inclusion of spin-orbit coupling. Due to smallness of this effect, the induced mass can modify the effective dipole expression as
\begin{equation}\label{eq:Phi}
\Phi(\vp{k},t)\approx\frac{e}{\hbar}\frac{\vp{E}\cdot\A{\varphi}_{\vp{k}}}{\sqrt{k^2+(\delta k)^2}}
\end{equation}
where $\hbar v_F\delta k=(\Delta_{so}-\lambda_R)$. The theory developed here serves as the starting point to analyze the optical response of graphene for an arbitrary order of interaction.

\section{Solution to the semiconductor bloch equations\label{section:5}}
As discussed, the SBEs in their original form describe the quasiclassical transport and interband excitation problems simultaneously. To convert the dynamical equations into a more convenient form, we proceed to decouple the transport and interband evolutions. It is noted that, neglecting the right side of SBE's, the Boltzman-type transport equation introduces a moving frame in the reciprocal space responsible for intraband evolution. It is worth noting that for a moderate applied field strength, the time evolution due to this moving frame can be considered as an adiabatic evolution in comparison to the 
interband one. Following this adiabatic argument, we now proceed to decouple the theory by introducing a moving frame in the reciprocal space. Assume that $\vp{k}_0(t)$ defines the \textit{primed} frame as
\begin{equation}\label{eq:k0}
\vp{k}_0(t)=-\frac{e}{\hbar}\int_{-\infty}^{t}\vp{E}\mathrm{d}t
\end{equation}
then $\vp{k}=\vp{k}_0+\vp{k}'$. The equation of motion in the primed frame is merely that of a simple two level problem

\begin{equation}\label{eq:primed-GBE}
\left\{
\begin{array}{lcl}
\pd{\mathcal{N}(\vp{k}'+\vp{k}_0,t)}{t} &=& -2{\Phi}(\vp{k}'+\vp{k}_0,t)\mathrm{Im}\left\{\mathcal{P}(\vp{k}'+\vp{k}_0,t)\right\}\\
&&
\\
\pd{\mathcal{P}(\vp{k}'+\vp{k}_0,t)}{t} &=& +i\Omega(\vp{k}'+\vp{k}_0)\mathcal{P}(\vp{k}'+\vp{k}_0,t)\\
&& +\frac{i}{2}{\Phi}(\vp{k}'+\vp{k}_0,t)\mathcal{N}(\vp{k}'+\vp{k}_0,t)
\end{array}\right.
\end{equation} 
These equations resemble the optical Bloch equations for a generic two level problem \cite{boyd2003nonlinear}. In
compliance with the standard notation of Bloch equations, we introduce $w_{\vp{k}}$, $u_{\vp{k}}$, $v_{\vp{k}}$ as 
 \begin{eqnarray}
w_{\vp{k}'}(t) &= &\mathcal{N}\left(\vp{k}'+\vp{k}_0(t),t\right)\label{eq:w}\\
u_{\vp{k}'}(t) &= & 2\mathrm{Re}\left\{\mathcal{P}\left(\vp{k}'+\vp{k}_0(t),t\right)\right\}\label{eq:u}\\
v_{\vp{k}'}(t) &= &-2\mathrm{Im}\left\{\mathcal{P}\left(\vp{k}'+\vp{k}_0(t),t\right)\right\}\label{eq:v}
\end{eqnarray} 
$w_{\vp{k}}$ is population inversion in the moving frame. In Eqs.~\eqref{eq:u} and \eqref{eq:v}, the factor of two and the minus sign are used to comply with
convention. $\xi_{\vp{k}}$ and $\omega_{\vp{k}} $ are also defined for mathematical convenience:
\begin{eqnarray}
\xi_{\vp{k}'}(t) &=& {\Phi}(\vp{k}'+\vp{k}_0(t),t)\label{eq:xi}\\
\omega_{\vp{k}'}(t) &=& \Omega (\vp{k}'+\vp{k}_0(t))
\end{eqnarray} 
The functions $\xi_{\vp{k}}$ and $\omega_{\vp{k}}$ are the analytical functions of the exciting field. According to Eq.~\eqref{eq:Phi}, $\xi_{\vp{k}}$ is the equivalent dipole in the moving frame. This dipole explicitly depends on the exciting field in the numerator of Eq.~\eqref{eq:Phi} and higher order nonlinear terms also exist due to motion of the primed frame. 
In adiabatic approximation the dynamical equations (not the excitations) are not directly affected  by the moving frame and therefore the time variations of $\omega_{\vp{k}}$ can be neglected as long as the pump wave intensity is not so large that multiphoton excitations take place.

The equations \eqref{eq:primed-GBE} provide an adequate description of resonant interband optical processes under conditions where relaxation processes can be neglected. Like the other two-level problems, the consideration of the relaxation processes, typically due to spontaneous emission, is of central importance \cite{boyd2003nonlinear}. We assume that the upper energy level decays to the lower energy level at a rate of $\gamma_1$. We also assume that the coherence terms decay with the dephasing rate of $\gamma_2$. Due to the strong $\vp{k}$ dependence of the transition dipole in the reciprocal space, the decay rates are expected to be $\vp{k}$ dependent. However, in the simplest approach, these parameters are assumed to be constant. The coupled Bloch equations can be converted to the well known optical Bloch equations in two level approximation \cite{boyd2003nonlinear}. From now on, we drop the prime in $uvw$-coordinate system.
\begin{align}
&\dot{w}_{\vp{k}}=-\gamma_1\left(w_{\vp{k}}-w^{eq}_{\vp{k}}\right)+\xi_{\vp{k}} v_{\vp{k}}\label{Wk}\\
&\dot{u}_{\vp{k}}=\omega_{\vp{k}}v_{\vp{k}}-\gamma_2 u_{\vp{k}}\\
&\dot{v}_{\vp{k}}=-\omega_{\vp{k}}u_{\vp{k}}-\gamma_2 v_{\vp{k}}-\xi_{\vp{k}} w_{\vp{k}}
\end{align}   
Where `dot'  denotes time derivative. The function $w^{eq}_{\vp{k}}$  is the population difference at equilibrium:
\begin{equation}
w^{eq}_{\vp{k}}=f(\mathcal{E}(\vp{k}))-f(-\mathcal{E}(\vp{k}))
\end{equation}
For a weak pump field, the inversion $w_{\vp{k}}$ tends to relax to $w^{eq}_{\vp{k}}$. The coherent terms, on the other hand, are the oscilatory functions of the field. To proceed further, the current response in the reciprocal space must be identified. According to  Eq.~\eqref{eq:rho100} together with Eq.~\eqref{eq:J100} the induced current is
\begin{multline}\label{eq:J1000}
\vp{J}=-2ev_F\sum_{\vp{k}}(\A{\vp{k}}\A{\vp{k}}+\A{\varphi}_{\vp{k}}\A{\varphi}_{\vp{k}}).\vec{m}_{\vp{k}}\\
=ev_F\sum_{\vp{k}}\left[-w_{\vp{k}-\vp{k}_0}\A{\vp{k}}+v_{\vp{k}-\vp{k}_0}\A{\varphi}_{\vp{k}}\right]
\end{multline}
Therefore the equation of motion describing the time evolution of $v_{\vp{k}}$  provides enough information to model the interband response of graphene. Neglecting the time variations of $\omega_{\vp{k}}$ in adiabatic approximation:
\begin{equation}
\ddot{v}_{\vp{k}}+2\gamma_2\dot{v}_{\vp{k}}+\left(\omega^2_{\vp{k}}+\gamma_2^2\right)v_{\vp{k}}=-\gamma_2\xi _{\vp{k}}w_{\vp{k}}-\dot{\xi}_{\vp{k}}w_{\vp{k}}-\xi_{\vp{k}}\dot{w}_{\vp{k}} 
\end{equation} 
Since $\omega^2_{\vp{k}}$ is much larger than $\gamma_2^2$ , we can drop $\gamma_2^2 v_{\vp{k}}$ to obtain the result:
\begin{equation}\label{master}
\ddot{v}_{\vp{k}}+2\gamma_2\dot{v}_{\vp{k}}+\omega^2_{\vp{k}}v_{\vp{k}}=-\gamma_2\xi _{\vp{k}}w_{\vp{k}}-\dot{\xi}_{\vp{k}}w_{\vp{k}}-\xi_{\vp{k}}\dot{w}_{\vp{k}} 
\end{equation}

Eq.~\eqref{master} describes a driven damped harmonic oscillator problem. The master equation \eqref{master} in conjunction with Eq.~\eqref{Wk} describe all linear and nonlinear properties of graphene. The origin of the nonlinear interband response in the moving frame lies in the fact that the coupling to the optical field depends parametrically on the inversion $w_{\vp{k}}$. Inversion is driven by the field stength $\xi_{\vp{k}}$ as described by Eq.~\eqref{Wk} which leads to a pure interband nonlinearity.
This set of equations can be solved iteratively. Expanding $v_{\vp{k}}$ and $w_{\vp{k}}$ into the powers of the exciting field, i.e. $\xi_{\vp{k}}$, gives a infinite series that contains the odd powers for $v_{\vp{k}}$ and even powers of the field for $w_{\vp{k}}$.
\begin{align}
&w_{\vp{k}}=w^{eq}_{\vp{k}}+\sum^{\infty}_{n=1}W_{\vp{k}}^{(2n)}\xi_{\vp{k}}^{2n}\label{eq:1000}\\
&v_{\vp{k}}=\sum^{\infty}_{n=1}V_{\vp{k}}^{(2n-1)}\xi_{\vp{k}}^{2n-1}\label{eq:1001}
\end{align}
In the rest of the paper, the $n$'th order expansion terms $w_{\vp{k}}^{(n)}$ and $v_{\vp{k}}^{(n)}$ are defined via
\begin{equation*}
w_{\vp{k}}^{(n)}=W_{\vp{k}}^{(n)}\xi_{\vp{k}}^{n}\quad,\quad v_{\vp{k}}^{(n)}=V_{\vp{k}}^{(n)}\xi_{\vp{k}}^{n} 
\end{equation*}

In addition to the pure interband multiphoton  process described above, a part of nonlinearity originates from the quasiclassical transport or intraband transitions. 
As will be clarified further in section \ref{section:7}, the frequency mixing effects in graphene arise from the pure intraband, pure interband and interband-intraband transitions. The pure intraband response occurs just because of the change in the the population difference. Using \eqref{eq:J1000}, the intraband contribution of the current is
\begin{equation}
\vp{J}_{intra}
=-ev_F\sum_{\vp{k}}w^{eq}_{\vp{k}-\vp{k}_0}\A{\vp{k}}
\end{equation}
then, the $n$'th order nonlinearity due to \textit{pure intraband process} can be obtained using Taylor expansion:
\begin{equation}\label{Master-Intraband}
\vp{J}^{(n)}_{intra}=(-1)^{n+1}\frac{ev_F}{n!}\sum_{\vp{k}}\A{\vp{k}}\left[\vp{k}_0(t).\nabla_{\vp{k}}\right]^n w^{eq}_{\vp{k}}
\end{equation}
An intuitive symmetry argument shows that in graphene as a centrosymmetric crystal, even orders of nonlinearity do not exist. In the succeeding sections the derivation of linear and third order conductivity of graphene is discussed.

\section{Linear Optical Response of Graphene\label{section:6}}
\subsection{Intraband linear response}
Eq.~\eqref{Master-Intraband} for $n=1$ gives the intraband conductivity tensor in the $k$-space
\begin{equation}
\OO{\sigma}^{(1)}_{intra}(\vp{k},\omega_p)=\frac{e^2v_F}{i\hbar\omega_p}\pd{w^{eq}_{\vp{k}}}{k}
\A{\vp{k}}\A{\vp{k}}\end{equation}
where we have assumed that the electric field is $\vp{E}=\T{\vp{E}}_p\exp(i\omega_p t) $ and $ \OO{\sigma}^{(1)}_{intra}({\vp{k}},\omega_p)$ is defined via 
\begin{equation*}
\vp{J}_{\vp{k}}=\OO{\sigma}^{(1)}_{intra}({\vp{k}},\omega_p)\cdot\vp{E}
\end{equation*}
Again the effect of radiation loss and collision in transport equation can be crudely incorporated in the equations in a phenomenological level.  We  assign an imaginary part to the frequency, i.e. the frequency $\omega_p$ will be replaced by $\omega_p-i\Gamma$.  This imaginary portion can be determined by fitting the analytical results to experimentally measured data. 
The main disadvantage of this method is, of course, certain lack of rigour, the method is, however less laborious.  
Performing integration in the reciprocal space, the off-diagonal terms vanish, we arrive at
\begin{multline}
\sigma^{(1)}_{intra}(\omega_p)=\frac{g_sg_v}{4\pi}\frac{e^2}{\hbar^2}\frac{1}{\left(i\omega_p+\Gamma\right)}\\
\int_0^{+\infty}\mathrm{d}\mathcal{E}\mathcal{E}\left[\pd{f(\mathcal{E})}{\mathcal{E}}-\pd{f(-\mathcal{E})}{\mathcal{E}}\right]
\end{multline}
where $g_s$ and $g_v$ are spin and valley degeneracy factors respectively. This equation can be simplified to a close form expression for the linear intraband conductivity
\begin{multline}\label{eq:intraband-linear}
\sigma^{(1)}_{intra}(\omega_p)=\frac{e^2}{\hbar}\frac{g_sg_v}{4\pi}\frac{k_BT}{\hbar(i\omega_p+\Gamma)}\\
\left[\frac{\mu}{k_BT}+2\ln\left(1+e^{-\mu/K_BT}\right)\right]
\end{multline}
\subsection{Interband linear response}
Interband linear optical response of graphene can be obtained using the master equation \eqref{master}. Linear optical response is a single photon process and unlike higher order terms it can be  obtained independetly for interband and intraband contributions. The first order solution of the Eq.~\eqref{master} can be derived by replacing $w_{\vp{k}}$ and $\dot{w}_{\vp{k}}$  with $w^{eq}_{\vp{k}}$ and $0$ respectively
\begin{multline}
\OO{\sigma}_{inter}(\vp{k},\omega_p)=\\
\frac{e^2}{\hbar}\frac{v_F}{\sqrt{k^2+(\delta k)^2}}\frac{\left(\gamma_2+i\omega_p\right)w^{eq}_\vp{k}}{\omega_p^2-2i \gamma_2\omega_p-\Omega^2_{\vp{k}}}\A{\varphi}_{\vp{k}}\A{\varphi}_{\vp{k}}
\end{multline} 
Integration over the reciprocal space and including the density of states gives
\begin{multline}\label{eq:interband-linear}
\sigma^{(1)}_{inter}(\omega_p)=\frac{e^2}{\hbar}\frac{g_sg_v}{4\pi}
\int_0^{+\infty}\mathrm{d}\mathcal{E}\frac{\mathcal{E}}{\sqrt{\mathcal{E}^2+(\Delta_{so}-
\lambda_{R})^2}}\\
\frac{\left(\gamma_2+i\omega_p\right)}{\omega_p^2-2i \gamma_2\omega_p-\Omega_{\vp{k}}^2}\left[f(\mathcal{E})-f(-\mathcal{E})\right]
\end{multline}
It is quite obvious  that in the linear regime the effect of spin orbit coupling can be neglected as the density of states on the energy axis linearly goes to zero and it removes the singularity and thus ${\mathcal{E}}/{\sqrt{\mathcal{E}^2+(\Delta_{so}-\lambda_{R})^2}}$ can be safely replaced by unity.  

Eqs. \eqref{eq:intraband-linear} and \eqref{eq:interband-linear} are identical to the results obtained from linear response theory and Kubo formulation \cite{gusynin2007magneto}. 


\section{Third Order Frequency Mixing in Graphene\label{section:7}}
Graphene as a  centrosymmetric  crystal does not exhibit second order nonlinearity and therefore the first nonlinear term is the third order one. In fact, to induce an optically biased second order response, translation symmetry must be broken. From what has been discussed so far, we know that the $\varphi_{\vp{k}}$ dependence of the $n$'th order optical conductivity tensor  in the reciprocal space, i.e.  $\OO{\sigma}^{(n)}_{\vp{k}}$, appears as 
$\OO{T}_{n}(\varphi_\vp{k})=\A{\alpha}_1\A{\alpha}_2\cdots\A{\alpha}_{n+1}$ where $\A{\alpha}_i$ can be either $\A{\vp{k}}$ or $\A{\varphi}_{\vp{k}}$. It is straightforward  to show that $\int_{0}^{2\pi}\OO{T}_n(\varphi_{\vp{k}})\mathrm{d}\varphi_{\vp{k}}$ vanishes when $n$ is an even integer.

Throughout this section we assume that three complex fields with the time dependence of $e^{i\omega_pt}$, $e^{i\omega_qt}$ and $e^{i\omega_rt}$ are mixing through the third order conductivity of graphene. As mentioned earlier, the third order optical response can be interpreted as a three-photon process and three different terms may contribute in the third order conductivity tensor: pure intraband term $\OO{\sigma}^{(3)}_{intra}(\omega_r,\omega_q,\omega_p)$, pure interband term $\OO{\sigma}^{(3)}_{inter}(\omega_r,\omega_q,\omega_p)$ and combination of the both $\OO{\sigma}^{(3)}_{intra-inter}(\omega_r,\omega_q,\omega_p)$.
\subsection{Intraband third order nonlinearity}
The third order intraband optical conductivity can be obtained using Eq.~\eqref{Master-Intraband}. Like the linear response term, an imaginary part would be assigned to the frequency components to include the effect of loss and collision. Exploiting the $\varphi_{\vp{k}}$ symmetry of $w^{eq}_{\vp{k}}$, the leading term in the third order nonlinearity term is given by
\begin{multline}\label{eq:intraband-3rd}
\OO{\sigma}^{(3)}_{intra}(\vp{k},\omega_r |\omega_q |\omega_p)=\\
\frac{e^4}{3!\hbar^3}\frac{v_F}{(i\omega_p+\Gamma)(i\omega_q+\Gamma)(i\omega_r+\Gamma)}\frac{\partial^3 w^{eq}_{\vp{k}}}{\partial k^3}\A{\vp{k}}\A{\vp{k}}\A{\vp{k}}\A{\vp{k}}
\end{multline}
The notation $(\dots,\omega_r |\omega_q |\omega_p )$ explicitly shows that the photons are ordered. This expression should be recognized from the third order intraband optical conductivity $\OO{\sigma}^{(3)}_{intra}(\vp{k},\omega_r ,\omega_q,\omega_p)$ which includes all possible permutations of the incoming photons. The formal integration of \eqref{eq:intraband-3rd} in the reciprocal space gives the third order intrband conductivity tensor
\begin{widetext}
\begin{equation}\label{eq:intraband-3rd-final}
\OO{\sigma}^{(3)}_{intra}\left(\omega_r,\omega_q,\omega_p\right)=\left(\frac{e^2}{\hbar}\frac{1}{E_c^2}\right)\frac{g_sg_v}{48\pi}\left(3\OO{T}_d+\OO{T}_{o}\right)
\mathcal{P}_I\left\{\frac{(k_BT)^3}{\hbar^3(i\omega_p+\Gamma)(i\omega_q+\Gamma)(i\omega_r+\Gamma)}
\frac{\exp\left(-\frac{\mu}{k_BT}\right)}{\left[1+\exp\left(-\frac{\mu}{k_BT}\right)\right]^2}\right\}
\end{equation}
\end{widetext}
Here we have made use of the intrinsic permutation operator $\mathcal{P}_I$, whose
meaning is that the right of it is to be summed over all possible
permutations of the input frequencies $\omega_p$, $\omega_q$ and $\omega_r$. The tensors $\OO{T}_d$ and $\OO{T}_{o}$ are defined as
\begin{equation}
\OO{T}_d=\A{\vp{x}}\A{\vp{x}}\A{\vp{x}}\A{\vp{x}}+\A{\vp{y}}\A{\vp{y}}\A{\vp{y}}\A{\vp{y}}
\end{equation}
\begin{multline}
\OO{T}_o=\\
\A{\vp{x}}\A{\vp{x}}\A{\vp{y}}\A{\vp{y}}+\A{\vp{y}}\A{\vp{y}}\A{\vp{x}}\A{\vp{x}}+\A{\vp{x}}\A{\vp{y}}\A{\vp{y}}\A{\vp{x}}+\A{\vp{y}}\A{\vp{x}}\A{\vp{x}}\A{\vp{y}}+\A{\vp{x}}\A{\vp{y}}\A{\vp{x}}\A{\vp{y}}+\A{\vp{y}}\A{\vp{x}}\A{\vp{y}}\A{\vp{x}}
\end{multline}
The parameter $E_c={(K_BT)^2}/{(e\hbar v_F)}$ is the characteristic electric field strength. In Eq.~\eqref{eq:intraband-3rd-final}, all terms inside the bracket are dimensionless and therefore 
$E_c$ gives a rough  estimate of the sufficient field stength to cause nonlinear effects. A glance shows that this coefficient is analogous to the Schwinger limit in QED giving the scale of the electric field that  Maxwell's equations are expected to become nonlinear. 

\subsection{Interband third order nonlinearity}
Pure interband third order nonlinearity can be obtained using the master equations \eqref{master} and \eqref{Wk} in the moving frame. Power expansion of the inversion and the polarization  in terms of the exciting field as Eqs.~\eqref{eq:1000} and \eqref{eq:1001} gives $v^{(3)}_{\vp{k}}$. Assuming that the first photon $\omega_p$ provides time variation for $v^{(1)}_{\vp{k}}(\omega_p)$, the first nonzero oscillatory component of the $w_{\vp{k}}$ as well as the third harmonics of $v^{(3)}_{\vp{k}}$ can be found from the following equations 
\begin{align}
&\dot{w}^{(2)}_{\vp{k}}+\gamma_1 w_{\vp{k}}^{(2)}=\xi_{\vp{k}}(\omega_q)v^{(1)}_{\vp{k}}(\omega_p)\\
&\nonumber
\\ 
&\ddot{v}^{(3)}_{\vp{k}}+2\gamma_2\dot{v}^{(3)}_{\vp{k}}+\omega^2_{\vp{k}}v^{(3)}_{\vp{k}} =\nonumber \\
 &\qquad -\gamma_2\xi _{\vp{k}}(\omega_r)w^{(2)}_{\vp{k}}-\dot{\xi}_{\vp{k}}(\omega_r)w^{(2)}_{\vp{k}}-\xi_{\vp{k}}(\omega_r)\dot{w}^{(2)}_{\vp{k}} 
\end{align}  
These equations can be solved simultaneously to find the third order interband conductivity tensor in the reciprocal space which is given in Eq.~\eqref{eq:sigma-3rd-k}. Performing integration in the reciprocal space and applying the permutation operator yields the final expression for the interband conductivity tensor. 
\begin{widetext}
\begin{multline}\label{eq:sigma-3rd-k}
\OO{\sigma}^{(3)}_{inter}(\vp{k},\omega_r |\omega_q |\omega_p) =-\frac{v_F e^4 }{(\hbar\sqrt{k^2+(\delta k)^2})^3}\\ 
\frac{\left[\gamma_2+i(\omega_p+\omega_q+\omega_r)\right]\left(\gamma_2+i\omega_p\right)w^{eq}_\vp{k}}{\left[(\omega_p+\omega_q+\omega_r)^2-2i\gamma_2(\omega_p+\omega_q+\omega_r)-\Omega^2_{\vp{k}}\right]\left[i(\omega_p+\omega_q)+\gamma_1\right]\left[\omega_p^2-2i \gamma_2\omega_p-\Omega^2_{\vp{k}}\right]} 
\A{\varphi}_{\vp{k}}\A{\varphi}_{\vp{k}}
\A{\varphi}_{\vp{k}}
\A{\varphi}_{\vp{k}}
\end{multline}  

\begin{multline}\label{eq:interband-3rd-final}
\OO{\sigma}^{(3)}_{inter}(\omega_r,\omega_q,\omega_p)=-\left(\frac{e^2}{\hbar}\frac{1}{E_c^2}\right)\frac{g_sg_v}{16\pi}\left(3\OO{T}_d+\OO{T}_{o}\right)
\mathcal{P}_I\left\{\frac{\hbar\left[\gamma_2+i(\omega_p+\omega_q+\omega_r)\right]\left(\gamma_2+i\omega_p\right)}{k_BT\left[i(\omega_p+\omega_q)+\gamma_1\right]}\right.
\\
\left.
\int_0^{+\infty}\frac{\mathrm{d}\mathcal{E}}{\mathcal{E}^2+(\Delta_{so}-\lambda_R)^2}
\frac{(k_BT)^5}{\hbar^4\left[(\omega_p+\omega_q+\omega_r)^2-2i\gamma_2(\omega_p+\omega_q+\omega_r)-\Omega^2_{\vp{k}}\right]\left[\omega_p^2-2i \gamma_2\omega_p-\Omega^2_{\vp{k}}\right]} \left[f(\mathcal{E})-f(-\mathcal{E})\right]\right\}
\end{multline}
\end{widetext}

The significant role of band gap opening and the band renormalization due to the spin-orbit coupling can be observed in Eq.~\eqref{eq:interband-3rd-final}. It is quite interesting that the integral given in Eq.~\eqref{eq:interband-3rd-final} possesses a first order singularity in the absence of spin-orbit coupling. 

\subsection{Intraband-Interband third order nonlinearity}
According to Eq.~\eqref{eq:1001}, the two level problem in the moving frame can provide the odd order of nonlinear response in terms of the effective dipole in the moving frame $\xi$. However, the motion of the frame can also be incorporated into the frequency mixing process. Through a second order intraband process the first and the second photons whose frequencies are $\omega_p$ and $\omega_q$ are contributing to the frequency mixing which accounts for the motion of the primed frame. Let us assume that, the third photon $\omega_r$ is absorbed through a first order interband process engendering $v^{(1)}_\vp{k}(\omega_p+\omega_q+\omega_r)$ . An intuitive argument shows that the first two photons mainly contribute in changing the population difference. The Taylor expansion of the ${w}_{\vp{k}}$ around the equilibrium state and plugging it into Eq.~\eqref{master} yields the third order conductivity tensor. The resultant equation is given in Eq.~\eqref{eq:inter-intra-k}. In Eq.~\eqref{eq:inter-intra-k} the higher order loss terms such as $\gamma_2\Gamma$ and $\Gamma^2$ have been neglected. Performing integration in the reciprocal space and applying the permutation operator yield the final expression for the third order conductivity tensor.

\begin{widetext}
\begin{multline}\label{eq:inter-intra-k}
\OO{\sigma}^{(3)}_{intra-inter}(\vp{k},\omega_r|\omega_q|\omega_p)=\\
-\frac{e^4v_F}{2\hbar^3}\frac{\gamma_2+2\Gamma+i(\omega_p+\omega_q+\omega_r)}{(i\omega_p+\Gamma)(i\omega_q+\Gamma)k}\pdt{w^{eq}_{\vp{k}}}{k}
\frac{\A{\varphi}_{\vp{k}}\A{\varphi}_{\vp{k}}\A{\vp{k}}\A{\vp{k}}}{(\omega_p+\omega_q+\omega_r)^2-2i\gamma_1(\omega_p+\omega_q+\omega_r)-\Omega^2_{\vp{k}}}
\end{multline}

\begin{multline}\label{eq:inter-intra}
\OO{\sigma}^{(3)}_{intra-inter}(\omega_r,\omega_q,\omega_p)=\left(\frac{e^2}{\hbar}\frac{1}{E_c^2}\right)\OO{T}_c\frac{g_sg_v}{32\pi}
\frac{(\gamma_2+2\Gamma)+i(\omega_p+\omega_q+\omega_r)}{\hbar(\omega_p-i\Gamma)(\omega_q-i\Gamma)}k_BT\\
\int_{0}^{+\infty}\mathrm{d}\mathcal{E} \pdss{\mathcal{E}}\left[f(\mathcal{E})-f(-\mathcal{E})\right]
\frac{(k_BT)^3}{(\omega_p+\omega_q+\omega_r)^2-2i\gamma_1 (\omega_p+\omega_q+\omega_r)-\Omega^2_{\vp{k}}}
\end{multline}
\end{widetext}

 where the tensor $\OO{T}_c$ is defined as
\begin{multline}
\OO{T}_c=3(\A{\vp{x}}\A{\vp{x}}\A{\vp{y}}\A{\vp{y}}+\A{\vp{y}}\A{\vp{y}}\A{\vp{x}}\A{\vp{x}})
+\A{\vp{x}}\A{\vp{x}}\A{\vp{x}}\A{\vp{x}}\\
+\A{\vp{y}}\A{\vp{y}}\A{\vp{y}}\A{\vp{y}}
-\A{\vp{x}}\A{\vp{y}}\A{\vp{x}}\A{\vp{y}}-\A{\vp{y}}\A{\vp{x}}\A{\vp{y}}\A{\vp{x}}-\A{\vp{y}}\A{\vp{x}}\A{\vp{x}}\A{\vp{y}}-\A{\vp{x}}\A{\vp{y}}\A{\vp{y}}\A{\vp{x}}
\end{multline}



\section{Results\label{section:8}}

The nonlinear optical behavior of graphene has been experimentally investigated by several groups \cite{hong2013optical,hendry2010coherent,zhang2012z}.  Four-wave mixing experiment \cite{ hendry2010coherent} and  Z-scan technique \cite{ zhang2012z}  have been among the most common methods to measure the nonlinear response of graphene. The results of the experiments confirm that graphene has an exceptionally high third-order susceptibility over a wide range of frequency.
 Depending on the measurement method and the sample quality, various research groups have reported different values for the bulk susceptibility and nonlinear refractive index of graphene. Despite the discrepancies between the different measurement results, all reliable experiments unanimously demonstrate that the nonlinear response of graphene is several orders of magnitude stronger than that of  all known semiconductors \cite{cheng2014third}.

\begin{figure}[!h]
\centering
\includegraphics[scale=0.23]{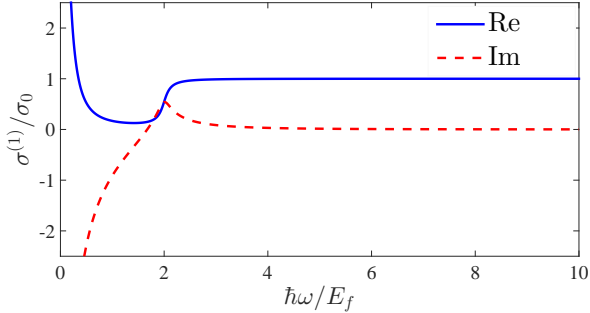}
\caption{Linear optical conductivity of graphene for normally incident plane wave. The parameter $\sigma_0=e^2/(4\hbar)$ is the universal optical conductivity.}
\label{fig-linear}
\end{figure}

\begin{figure*}[!ht]
\centering
\begin{subfigure}[]{
\includegraphics[width=0.45\linewidth]{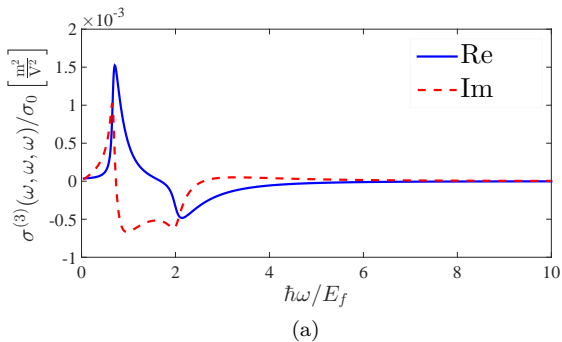} 
\label{fig:chi3_third} }
\end{subfigure}
\hfill
\begin{subfigure}[]{
\includegraphics[width=0.45\linewidth]{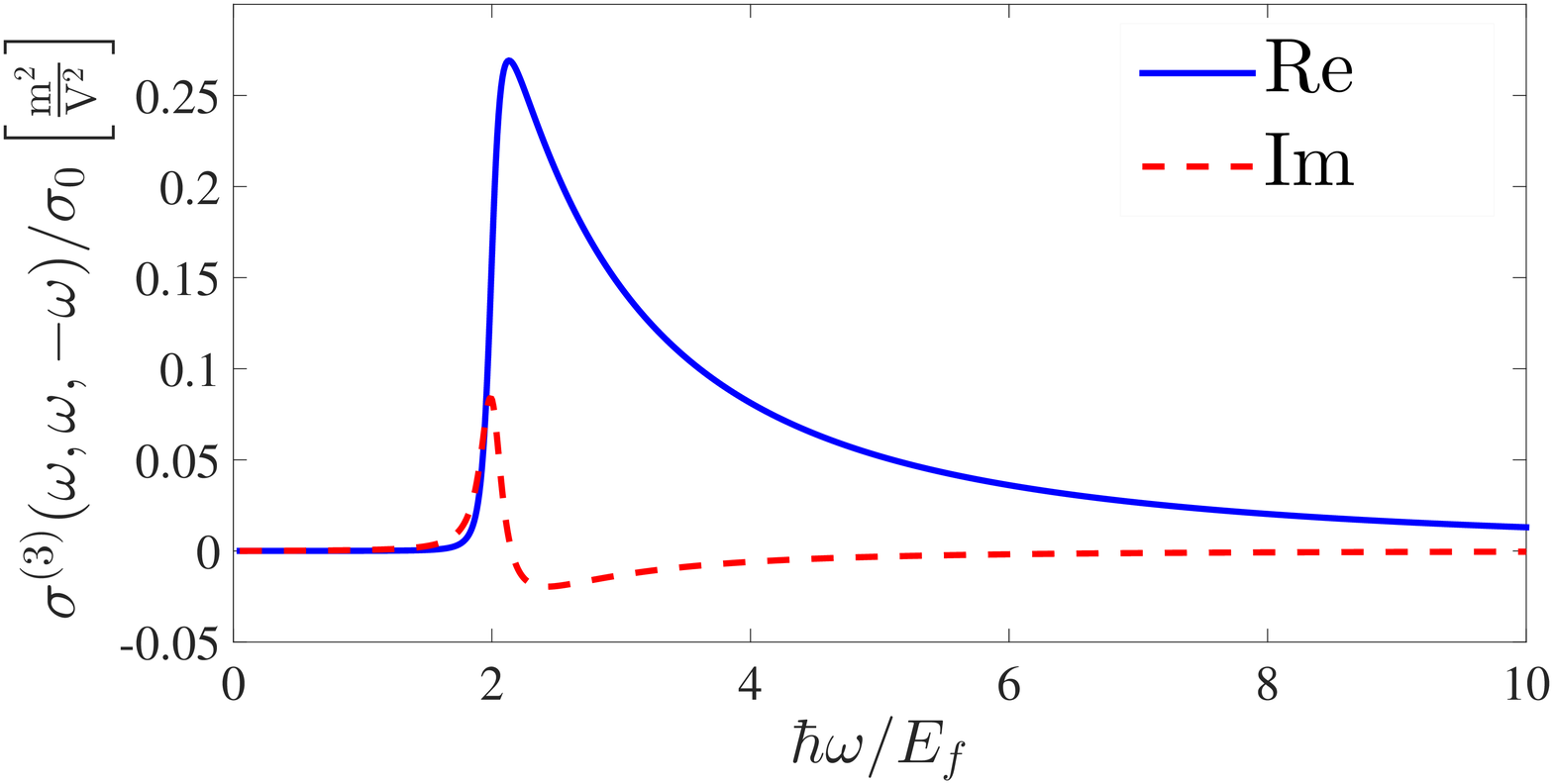} 
\label{fig:chi3_kerr} }%
\end{subfigure}

\caption{The third order conductivity of graphene normazlied to  $\sigma_0=e^2/(4\hbar)$  in the unit of  $\mathrm{\frac{m^2}{V^2}}$  (a) $\sigma^{(3)}_{xxxx}(\omega,\omega,\omega)$ and (b)  $\sigma^{(3)}_{xxxx}(\omega,\omega,-\omega)$.}
\label{fig-chi3}
\end{figure*}

The theoretical predictions for the linear and nonlinear optical response of graphene are shown in Figs.\ref{fig-linear} and \ref{fig-chi3}. In our calculations, the unknown parameters are selected according to the experimental results. The phenomenological intraband scattering rate $\Gamma$ and two-level interband relaxation coefficients ($\gamma_1$ and $\gamma_2$) obviously depend on the frequency, temperature and quality of the sample. A full theoretical investigation of the Drude conductivity of graphene and possible origins of the relaxation coefficients is given in \cite{stauber2007electronic}. According to the experimental results reported in \cite{horng2011drude,malard2013observation}, the relaxation coefficients are typically around tens of meV ($\Gamma,\gamma_{1,2}\sim 10\mathrm{meV}$).
To highlight the resonant features of the linear and nonlinear response, it is assumed that graphene is held at the temperature of $T\approx 0\mathrm{K}$. It is also assumed that the Fermi energy level is around $100\mathrm{meV}$. The energy of the incoming photon(s) are normalized to the Fermi energy.

The linear optical conductivity of graphene is shown in Fig.~\ref{fig-linear}. It is observed that the optical absorption of graphene is universal and independent of the frequency for the photon energies of $\hbar\omega >\sim 2E_f$. The universal behavior of the optical conductivity in graphene can be explained by the two-dimensionality and the invariance of the condensed matter system \cite{bacsi2013,Hatsugai2013}. However, the dispersionless character of the absorption spectral is not topologically protected and even the inclusion of the higher order terms -such as triangular warping \cite{katsnelson}- can deviate it from the universally flat response. 

Fig.~\ref{fig-chi3}  shows the frequency dependence of the third order conductivity of graphene.  
It can be shown that the third order conductivity tensor is mainly dominated by the interband transitions and Rabi-type oscillations determine their frequency dependence for high-energy photons. The frequency variation of  the $\sigma^{(3)}_{xxxx}(\omega,\omega,\omega)$  which is responsible for the third harmonic generation is shown in Fig.~\ref{fig:chi3_third}. The resonant features of this response can be explained based on the Rabi oscillations in the reciprocal space. The integrand in Eq.~\eqref{eq:interband-3rd-final} for $\omega_{p,q,r}=\omega$ possesses four simple poles at $\Omega_\vp{k}\approx \pm \omega,\pm \omega/3$. In the absence of Pauli blocking , i.e. for $\hbar\Omega_{\vp{k}}>\sim 2E_f$, the interband transitions take place. The overall response is the superposition of the broadened resonances in the reciprocal space. The resonances around the $\hbar\omega\sim 2E_f$ are stronger leading to appearing of the peaks around $\hbar\omega\sim \frac{2}{3}E_f$ and $\hbar\omega\sim 2E_f$.

Our theoretical prediction of  $\sigma^{(3)}_{xxxx}(\omega,\omega,-\omega)$ is also plotted in Fig.~\ref{fig:chi3_kerr}. This part of the nonlinearity contributes in the nonlinear refractive index. This component of the nonlinear response exhibits resonant behavior around $\hbar\omega\sim 2E_f$.  The nature of this resonant behavior can be explained by looking at the imaginary of the integrand in Eq.~\eqref{eq:interband-3rd-final} . The integrand for $\omega_{p,q,r}=\omega,\omega,-\omega$ has two second order poles at $\Omega_{\vp{k}}\approx \pm \omega$ . The amplitude of the absorption for such poles is roughly proportional to the slop of population difference. Those resonances are significantly stronger around the Fermi energy level where the population difference abruptly changes. A physical interpretation of this behavior can be presented based on two photon absorption of a bunch of two level systems in the reciprocal space.


\begin{figure}[h]
\centering
\includegraphics[scale=0.23]{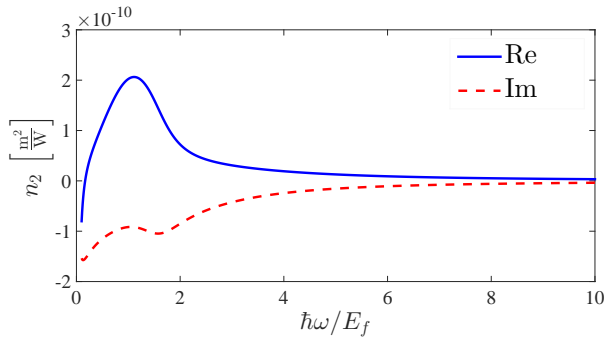}
\caption{The Kerr nonlinear coefficient of graphene ($n_2$ defined in Eq.~\eqref{eq:n2}) at $T=300\mathrm{K}$. Over a wide range of frequency the Kerr coefficient is around  $10^{-11}\mathrm{m^2W^{-1}}$.}
\label{fig-n2}
\end{figure}

The equivalent third order bulk susceptibility of graphene is related to the third order surface dynamic conductivity via
\begin{equation}
\chi^{(3)}(\omega_p,\omega_q,\omega_r)=\frac{\sigma^{(3)}_{xxxx}(\omega_p,\omega_q,\omega_r)}{i(\omega_p+\omega_q+\omega_r) d_{gr}\varepsilon_0}
\end{equation}
where $d_{gr}$ is the equivalent thickness of graphene which is typically around $d \approx 3\AA $ \cite{cheng2014third,hendry2010coherent} and $\varepsilon_0$ is the free space permittivity. 
Obviously for the case of graphene, the definition of the nonlinear bulk susceptibility is ambiguous due to the arbitrariness in the definition of the thickness of the two-dimensional structure. In the Kerr-type nonlinear response, the dependence of the complex refractive index $n$ on the intensity of light $I$  is given by 
\begin{equation}\label{eq:n2}
n=n_0+n_2I
\end{equation}
Where $I=2\epsilon_0 \mathrm{Re}\{n_0\}c\abs{\vp{E}}^2$ ($c$ is the speed of light). The nonlinear coefficient $n_2$ is related to the bulk susceptibility $\chi^{(3)}(\omega,\omega,-\omega)$ as \cite{delCoso2004}
\begin{equation}
\label{eq:kerr}
n_2=\frac{3}{4\epsilon_0 c\abs{n_0}^2}\chi^{(3)}(\omega,\omega,-\omega)\left[1-i\frac{\mathrm{Im}\{n_0\}}{\mathrm{Re}\{n_0\}}\right]
\end{equation}
It is easy to show that for the case of graphene this expression is merely independent of the particular choice of $d_{gr}$ and it introduces an intrinsic parameter. Fig.~\ref{fig-n2} displays our theoretical prediction for the real and imaginary parts of the Kerr coefficient $n_2$ at the room temperature.  

As the figure shows in a wide range of frequency and far from the resonances, the nonlinear coefficient is around $n_2\approx 10^{-11}\mathrm{m^2W^{-1}}$. This results is in a reasonably good agreement with the experimental results provided in \cite{zhang2012z}. 
This curve also indicates that the Kerr nonlinearity of graphene is much stronger than all known nonlinear semiconductors such as  $\mathrm{GaAs}$, $\mathrm{Ge}$ and $\mathrm{AlGa}$. According to the experimental results presented in Ref.~\cite{sheik1991dispersion} the nonlinear Kerr coefficient  for those materials is in the order of $10^{-16}\mathrm{m^2 W^{-1}}$ which is obviously much smaller than that of graphene. It is quite interesting that nonlinear Kerr coefficient can be tuned by changing the Fermi energy level and extremely high Kerr type nonlinearity can be achieved in appropriately gated graphene monolayer and chiral multilayer graphene.



\section{conclusion\label{section:conclusion}}
A semiclassical theory of light-graphene interaction in linear and nonlinear regimes has been detailed. Focusing on the scale-invariancy and chiral character of Bloch quasi-particles in the graphene Hamiltonian, the Semiconductor Bloch Equations have been formulated. The advantage of SBEs is two-fold: first they provide a convenient mathematical scheme leading to the analytical expressions for different contributions of the linear and the nonlinear optical response of graphene. Moreover, SBEs encode the topological properties of the band structure in an effective dipole expression appearing in the equations. The mathematical structure of this effective dipole reveals the distinct optical response of graphene. 

The exotic transport and optical properties of graphene can be attributed to the chirality of the charged carriers.  Giving the mathematical description of the chirality and employing SBEs, it has been shown that this chirality leads to a remarkably strong nonlinear optical response. 

Using SBEs, the problem of the interaction can then be decomposed into the quasicalssical Boltzman transport and the interband time evolution. The nonlinear parts of the optical response can be classified as pure intraband, pure interband and combination of the both. Introducing a novel mathematical framework, analytical expressions for different contributions of the conductivity tensors have been derived for the first time.  We have shown that a suitable band renormalization is required to remove a singularity in the third order interband optical response of graphene resulting form the chiral symmetry and the scale invarince of the band structure.

The third order susceptibility and nonlinear refractive index have been calculated. It is shown that  our prediction for the Kerr nonlinear  coefficient for graphene is in reasonably good agreement with the experimentally obtained results. It has been demonestrated that Kerr nonlinear coefficient of graphene can be tuned by changeing the Fermi energy level and significantly strong Kerr nonlinearity can be attained in a gated graphene monolayer.

\begin{acknowledgements}
 This work has been supported by Natural Science and Engineering Research Council of Canada (NSERC)
and BlackBerry (former Research in Motion). We would like to thank Prof. Anton Burkov for very fruitful discussions.
\end{acknowledgements}

\bibliography{BIB}

\end{document}